\documentclass[journal,12pt,onecolumn,draftclsnofoot,]{IEEEtran}
\usepackage[utf8x]{inputenc}

\usepackage{amsmath,amssymb,dsfont,stfloats,url}
\usepackage{color}
\usepackage[pdftex]{graphicx}
\usepackage{float}
\usepackage[caption = false]{subfig}
\usepackage{algpseudocode,algorithm,algorithmicx}
\usepackage{afterpage}
\usepackage{balance}
\usepackage{cite}
\usepackage{array}
\usepackage{verbatim}
\usepackage{comment}
\usepackage{mathtools}
\usepackage{multicol}

\usepackage{verbatim}

\usepackage{amsmath}
 
\usepackage{pstricks}

\newtheorem{example}{{\bf Example}}
\newfont{\bb}{msbm10 scaled 1100}

\newcommand{\RR}{\mbox{\bb R}}

\newcommand{\FF}{\mbox{\bb F}}

\newcommand{\dv}{{\bf d}}

\newcommand{\Bm}{{\bf B}}

\newcommand{\Zm}{{\bf Z}}

\newcommand{\Ac}{{\cal A}}
\newcommand{\Bc}{{\cal B}}

\newcommand{\Ec}{{\cal E}}
\newcommand{\Fc}{{\cal F}}

\newcommand{\Hc}{{\cal H}}

\newcommand{\Lc}{{\cal L}}

\newcommand{\Pc}{{\cal P}}

\newcommand{\Rc}{{\cal R}}
\newcommand{\Sc}{{\cal S}}
\newcommand{\Tc}{{\cal T}}
\newcommand{\Uc}{{\cal U}}

\newcommand{\Rsf}{{\mathsf R}}

\usepackage{hyperref}
\hypersetup{
    bookmarks=true,         % show bookmarks bar?
    unicode=false,          % non-Latin characters in Acrobat�s bookmarks
    pdftoolbar=true,        % show Acrobat�s toolbar?
    pdfmenubar=true,        % show Acrobat�s menu?
    pdffitwindow=false,     % window fit to page when opened
    pdfstartview={FitH},    % fits the width of the page to the window
    pdfnewwindow=true,      % links in new window
    colorlinks=true,       % false: boxed links; true: colored links
    linkcolor=red,          % color of internal links (change box color with linkbordercolor)
    citecolor=green,        % color of links to bibliography
    filecolor=blue,      % color of file links
    urlcolor=blue           % color of external links
}
\graphicspath{ {./Figures/} }
\DeclareGraphicsExtensions{.eps,.pdf,.png,.jpg,.gif,.jpeg,.pstex}

\makeatletter
\def\munderbar#1{\underline{\sbox\tw@{$#1$}\dp\tw@\z@\box\tw@}}
\makeatother

\algdef{SE}[SUBALG]{Indent}{EndIndent}{}{\algorithmicend\ }%
\algtext*{Indent}
\algtext*{EndIndent}

\begin{document}

\title{Coded Caching over Multicast Routing Networks}

\author{
Mozhgan~Bayat,~\IEEEmembership{Student Member,~IEEE,} 
Kai~Wan,~\IEEEmembership{Member,~IEEE,}
and~Giuseppe Caire,~\IEEEmembership{Fellow,~IEEE}
\thanks{
M. Bayat, K.~Wan, and G.~Caire are with the EECS Faculty, Technische Universit\"at Berlin, 10623 Berlin, Germany
(e-mail: bayat@tu-berlin.de; kai.wan@tu-berlin.de; caire@tu-berlin.de). 
This work is partially funded by the European Research Council under the ERC Advanced Grant N. 789190, CARENET.
A short version of this paper was presented in IEEE 20th International Workshop on Signal Processing Advances in Wireless Communications (SPAWC), 2019}

}

\maketitle

\vspace{-1cm}

\begin{abstract}
The coded caching scheme originally proposed by Maddah-Ali and Niesen (MAN) transmits coded multicast messages from a server 
to users equipped with caches via a capacitated shared-link and was shown to be information theoretically
optimal within a constant multiplicative factor.  This work extends the MAN scheme to a class of two-hop wired-wireless 
networks including one server connected via fronthaul links to a layer of $H$ helper nodes (access points/base stations), 
which in turns communicate via a wireless access network to $K$ users, each equipped with its own cache.
Two variants are considered, which differ in the modeling of the access segment. Both models should be regarded as {\em abstractions} 
at the ``network layer'' for physical scenarios such as local area networks and cellular networks, spatially distributed over a certain coverage area.  
The key focus of our approach consists of routing MAN-type multicast messages through the network and formulating the optimal routing 
scheme as an optimization problem that can be solved exactly or for which we give powerful heuristic algorithms.  
Our approach solves at once {\em many of the open practical problems} identified as stumbling blocks for the application of coded caching 
in practical scenarios, namely:  asynchronous streaming sessions, finite file size, scalability of the scheme to large and spatially distributed 
networks, user mobility and random activity (users joining and leaving 
the system at arbitrary times), decentralized prefetching of the cache contents, 
end-to-end encryption of HTTPS requests, 
which renders the helper nodes oblivious of the user demands. 
\end{abstract}

\begin{keywords}
Coded Caching, Multicast Routing, Reduced Subpacketization Order, Linear Programming.
\end{keywords}

%%%%%%%%%%%%%%%%%%%%%%%%%%%%%%%%%%%%%%%%%%%%%%%%%%
\section{Introduction}
\label{sec:Introduction}

Due to the growing consumption of on-demand multimedia content, a clever use of caching exploiting the low-cost storage capacity on user devices 
plays a key role in the design of efficient content distribution schemes. In this context, 
{\it caching} refers in general to the prefetching of popular files (or blocks thereof) at the edge nodes such that the traffic load can be reduced when users' demands are revealed. For example, \cite{Golrezaei2012femto} introduced {\em femtocaching} in wireless networks, where caching is performed at ``helper'' nodes modeling Access Points (APs) or Base Stations (BSs) in a spatially distributed  wireless local area or cellular network. More recently, Fog Radio Access Networks (F-RAN) have been proposed, where helpers may posses local caches as well as baseband processing units. 
By letting helpers store popular files in their cache memories, \cite{shamai2016opt}  treats the joint design of the centralized (cloud) and decentralized (fog) processing to satisfy users' demands. 

The above works, as well as many others that would be too long to mention here, are based on {\em uncoded caching}, where the demanded files are directly transmitted from the caches at the helpers and from the server through the network. 
A different line of works considers  {\em coded caching} strategies, where the cache content at the 
users is exploited as side information such that coded {\em multicast messages} are simultaneously  useful for many users.  
Through coding, individual user demands (unicast traffics) are converted into multicast messages which may be better suited to the broadcast property of the transmission medium.  The first coded caching scheme was originally proposed by Maddah Ali and Niesen (MAN) in \cite{maddah2014fundamental}.
In the MAN setting, the server has a library of $N$ files and broadcasts its transmission to $K$ users, each with a cache containing up to $M$ files, 
through an error-free shared-link.  The MAN scheme consists of two phases: {\it prefetching} and {\it delivery}. 
%During the prefetching phase, users store some segments of the files without knowledge of future demands. 
%During the delivery phase, each user requests one file and the server broadcasts some coded packets such that each user can recover its desired file with the %help 
%of its cache. 
The objective is to design the prefetching scheme (cache content) and a coded delivery scheme (formation of the coded multicast messages) such that
the worst-case load over all possible user demands is minimized. 
The MAN caching scheme  was proved in \cite{ontheoptimality} to be optimal  under the constraint of uncoded prefetching
 (i.e., each user directly stores a collection of segments of the library files in its cache) when $N \geq K$. 
By removing the redundant MAN multicast messages when $N < K$, \cite{yu2017exact} improved 
the delivery phase and proved exact optimality for any $N$ and $K$ under uncoded prefetching. In general, the MAN scheme
with the improvement of \cite{yu2017exact} was shown in \cite{yu2018characterizing} to be optimal within a factor of 2 over all possible schemes, 
even removing the uncoded prefetching condition.  

In the MAN scheme, the prefetching phase is centrally coordinated.
In practice, coordination may not be possible. 
For example, in a mobile network over an extended coverage area it would be impractical that all users receive from the same 
giant transmitter. In a typical wireless network scenario, the server communicates to the users via a layer of spatially distributed helper nodes (APs/BSs).  
Hence, due to mobility, the local cache configuration in each cell cannot be centrally pre-designed during the prefetching phase. 
Furthermore, users join and leave the network at arbitrary times and in an uncoordinated fashion. Therefore, {\em decentralized} prefetching schemes
are needed in practice. 

Another important aspect of practical systems is that user on-demand streaming sessions start and end at arbitrary times 
and are formed by sequences of HTTP requests, fetching sequentially chunks of the streamed video file \cite{HTTPvideo} and \cite{iDASH}.  
 The fundamental performance in this context is the {\em average delivery time} per chunk, which much be 
(slightly) smaller than the chunk playback time in order to keep the probability of empty playback buffer (buffer underrun) 
sufficiently small   \cite{WiFlix}. In order to handle the streaming sessions asynchronism in coded caching, 
each large video file can be divided into blocks, which are themselves identified with the 
``library files'' of the coded caching scheme \cite{decMAN}.  
In order to avoid a long waiting time before starting a streaming session, the block playback duration should not exceed 
a few tens of seconds.  For example, with a streaming rate of 2 Mbit/s, and 10s blocks, 
each effective file item in the library has size of 20 Mbits.  This means that even though the actual video files may be very large, 
the effective length of the library files treated by the coded caching scheme is limited. 
This imposes a limitation of the {\em subpacketization order} of the coded caching scheme, i.e.,
the number of segments in which each library file is divided in order to be cached in the prefetching phase.
In particular, in the MAN scheme the subpacketization order grows exponentially with the number of users $K$, which makes the scheme impractical for large networks.  
%In contrast, we notice that in the literature on coded caching a lot of attention has been dedicated to the case of $N < K$, i.e., less library files than users 
%(e.g., see~\cite{yu2017exact,multiJi2014}). 
%However, we argue here that in the context of 
%practical video streaming the case $N < K$ is basically irrelevant. 
%For example, consider a server with $1000$ video files, each of which is formed by $500$ chunks. Identifying the chunks as the library files we have  
%$N = 5 \times 10^5$. Consider a large network with 100 cells serving on average $10$ streaming users per cell, resulting in $K = 10^3$ users.
%In such scenarios, it can be expected that $N$ is larger than $K$ by more than 2 orders of magnitude. 
%Therefore, in the following we will assume $N \geq K$. 

Beyond user mobility and limited subpacketization order, there are several other practical issues that must be addressed in order to make the coded caching paradigm  suitable for practical implementations (see for example the discussion in  \cite{5gcaching}). 
In particular, here we would like to mention: i) the problem of HTTP encryption, for which the user request can be decrypted only by the 
server, which belongs to the content owner (e.g., Netflix, Google, Apple) and not  by intermediate helper nodes (APs and BSs belonging to some 
wireless network operator); ii) the fact that a content delivery scheme is typically run ``above IP'', i.e., at the application and transport layer, and does not involve the underlying  lower layers such as PHY and MAC, which follow some existing legacy standard (e.g., IEEE802.11, LTE, 5G NR) and are not under the control of the content distribution system;  iii) the fact that in modern content distribution networks a single content server handles very large regions corresponding to tens of cells (e.g., an entire large  metropolitan area) \cite{netflix}.
In particular, point (i) rules out the possibility of ``fog'' caching at the helper nodes (considered for example in~\cite{hierarchicalcaching,ji2015cachingasilomar,KaiFog}),  since the helpers are not supposed to store content and are oblivious of the user requests; 
point (ii) rules out the possibility of joint PHY and caching design, as for example
combining coded caching with MIMO zero-forcing precoding as in  \cite{lampiris2018adding}; 
 and  point (iii) greatly de-emphasize the relevance of various ``multiserver'' models  \cite{shariatpanahi2016multi,navid2017interference}.  

\begin{figure}[t]
 	\centering
 	\includegraphics[width=0.55\textwidth]{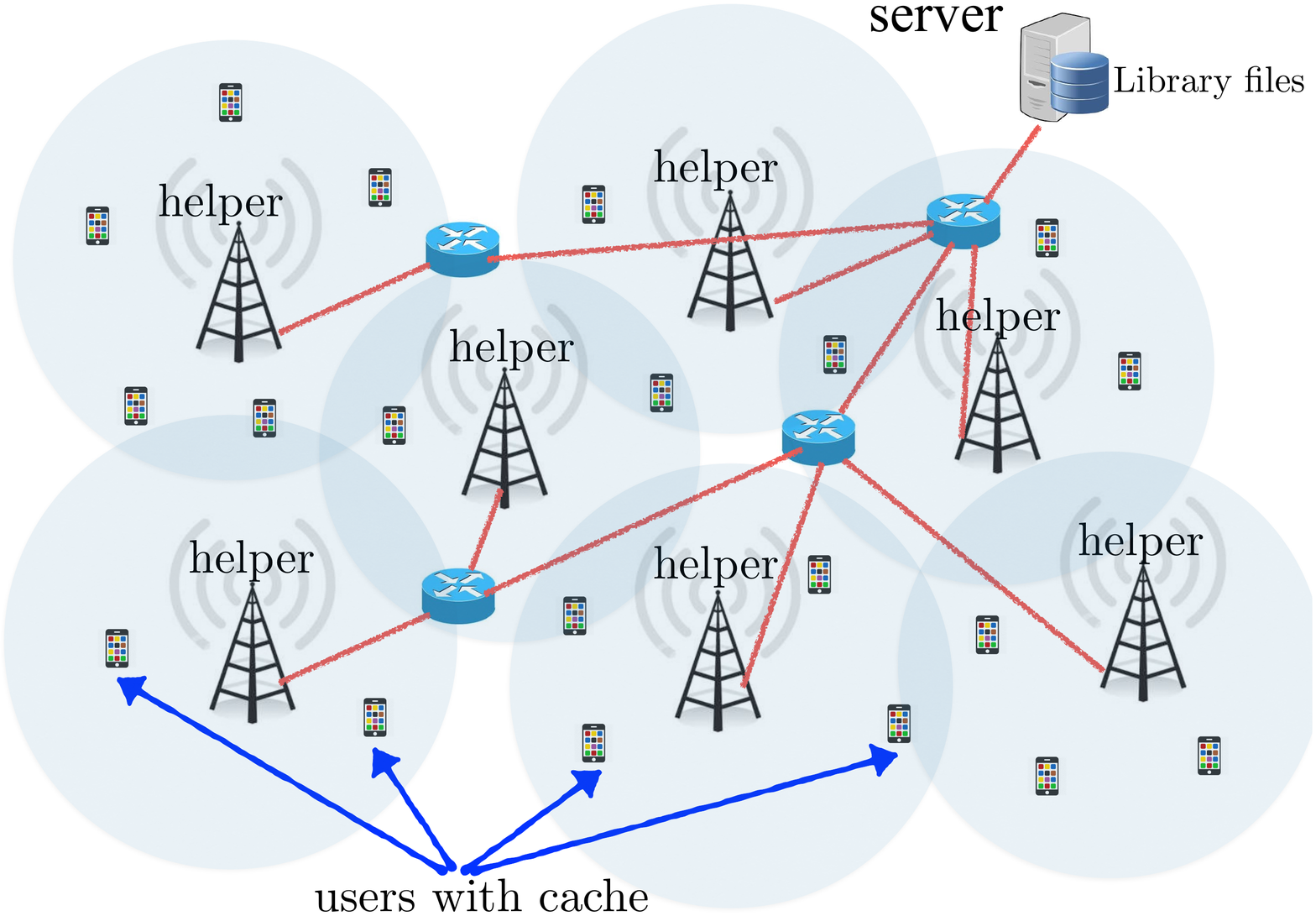}
 	\caption{A pictorial representation of the networks considered in this work, formed by one server communicating to $K$ users via $H$ helpers, 
	and routing coded caching packets via a multicast  IP network.}
 	\label{fig:network-concept}
 \end{figure}

Overall, these considerations motivate us to study network models resulting from the network-layer abstraction the underlying 
physical wireless network.  Our model consists of one server, co-located with a library of $N$ files, a layer of $H$ helper nodes, 
connected to the server via capacitated {\em fronthaul} links, and $K$ users, connected to a limited number of helpers via a wireless {\em access network},
depending on their geographic location on the network area.  
The server communicates to the users via the helper nodes, and uses a multicast routing protocol (typically IP multicast) 
in order to leverage the multicast nature of the coded caching delivery phase.  
The helper nodes represent APs/BSs and are oblivious of the user demands and of the caching scheme, 
i.e., they can only forward to the users what they receive from the server. 
This  capture the fact that helpers can only read the IP routing control of the packets, but cannot process, combine, or store such packets. 
Fig.~\ref{fig:network-concept} represents qualitatively the class of networks treated in this paper. 

%%%%%%%%%%%%%%%%%%%%%%%%%%%%%%%%%%%%%%%%%%%%%%%%%%%%%%%%%%%%%
\subsection{Contributions}

This work extends the MAN scheme to the aforementioned two-hop network by considering the following two variants:
i) A topological network formed by non-interfering links of limited capacity 
(referred to as ``topological network'' model in the following); 
ii)  A network with broadcast constraints at the helpers and collision interference at the users
(referred to as ``broadcast/collision network'' model in the following).  
{\em For both models, we propose schemes
that solve at once all the problems mentioned before, i.e., subpacketization order, large network scalability, decentralized/asynchronous users' activity, and
transparency of the helper nodes which must only execute a standard multicast routing scheme oblivious of the content files and user requests, 
and therefore can support HTTP encryption.}  Our main contributions are summarized as follows:

1) For the topological network model, we propose a novel scheme based on decentralized random cache replication prefetching and optimized routing of 
	the MAN-type coded multicast messages. The routing optimization can be solved via a sequence of linear programs (LPs), 
	with linear complexity in the network size.  For comparison, we consider also   a direct extension of the so-called {\em multiround delivery} 
	scheme in \cite{parrinello2018fundamental, jin2016}. The proposed novel routing strategy generally outperforms the extended multiround scheme since the latter is a particular feasible point of the optimization problem yielding the former. 

2) For the broadcast/collision model, we first propose a two-step baseline reuse scheme, where in the first step we assign to the same transmission resource (e.g., a time slot or a frequency subband)  groups of mutually non-interfering helpers  by graph coloring, and in the second step we assign the users to the helpers in an optimal way (which may be computationally hard) or using a greedy approach. 
Then, for the same broadcast/collision model, we propose a novel scheme that embraces interference and serves users as they become interference-free. 
This approach is nicknamed {\em avalanche scheme} since the users are freed of interference progressively as some helper finish serving its own list of interference-free users, and this property propagates through the network as an avalanche. The scheme has a practical appeal since it could be implemented via  
the CSMA protocol,  where collisions are discarded at the receivers (the users) and helpers pause transmission when they have finished serving the users in their service list. The avalanche scheme not only avoids the graph coloring problem which is generally NP-hard, but also outperforms the reuse scheme since instead of following a fixed reuse partition of the transmission resource makes use of it in a more adaptive and opportunistic way.
We show that for sparse graphs, which is the case for  spatially distributed wireless networks, the complexity of the avalanche scheme is  linear in the 
network size.  
% 	 \item Simulation results reveal that  the avalanche scheme is more robust and delivery time is only mildly dependent on 
%	 the number of distinct cache configurations in the network. Since the subpacketization order grows exponentially with the number of distinct cache 
%	 configurations, this means that we can operate the system with a pretty small number of distinct cache configurations (i.e., achieving a small %subpacketization order) and yet achieve a delivery time very close to the minimum (when the number of cache configurations is equal to the number of users). %This feature, combined with the proximity to a CSMA mechanism, makes the novel avalanche scheme very attractive in practice.

\paragraph*{\textbf{Notation Convention}}
%Calligraphic symbols denote sets, and  bold symbols denote vectors. We use $|\cdot|$ to represent the cardinality of a set or the length of a sequence. 
We define $[a:b]=\left\{ a,a+1,\ldots,b\right\}$, $[n] = [1:n]$, and
$\mathcal{A\setminus B}=\left\{ x\in\Ac : x\notin\Bc\right\}$. 
To denote a generic collection of indexed objects $X_j$ we use the notation $\{X_j\}$, where
the index range is clear from the context. The symbol $| \cdot |$ indicates the number of elements of a set or the length of a sequence appearing as argument 
inside the bars.

%%%%%%%%%%%%%%%%%%%%%%%%%%%%%%%%%%%%%%%%%%%%%%%%%%%%%
\section{Topological Networks of Non-Interfering Links}
\label{sec:Topological-networks}

\begin{figure}
    \centering
    \subfloat[Layout geometry and user-helper association.]{\includegraphics[width=6cm]{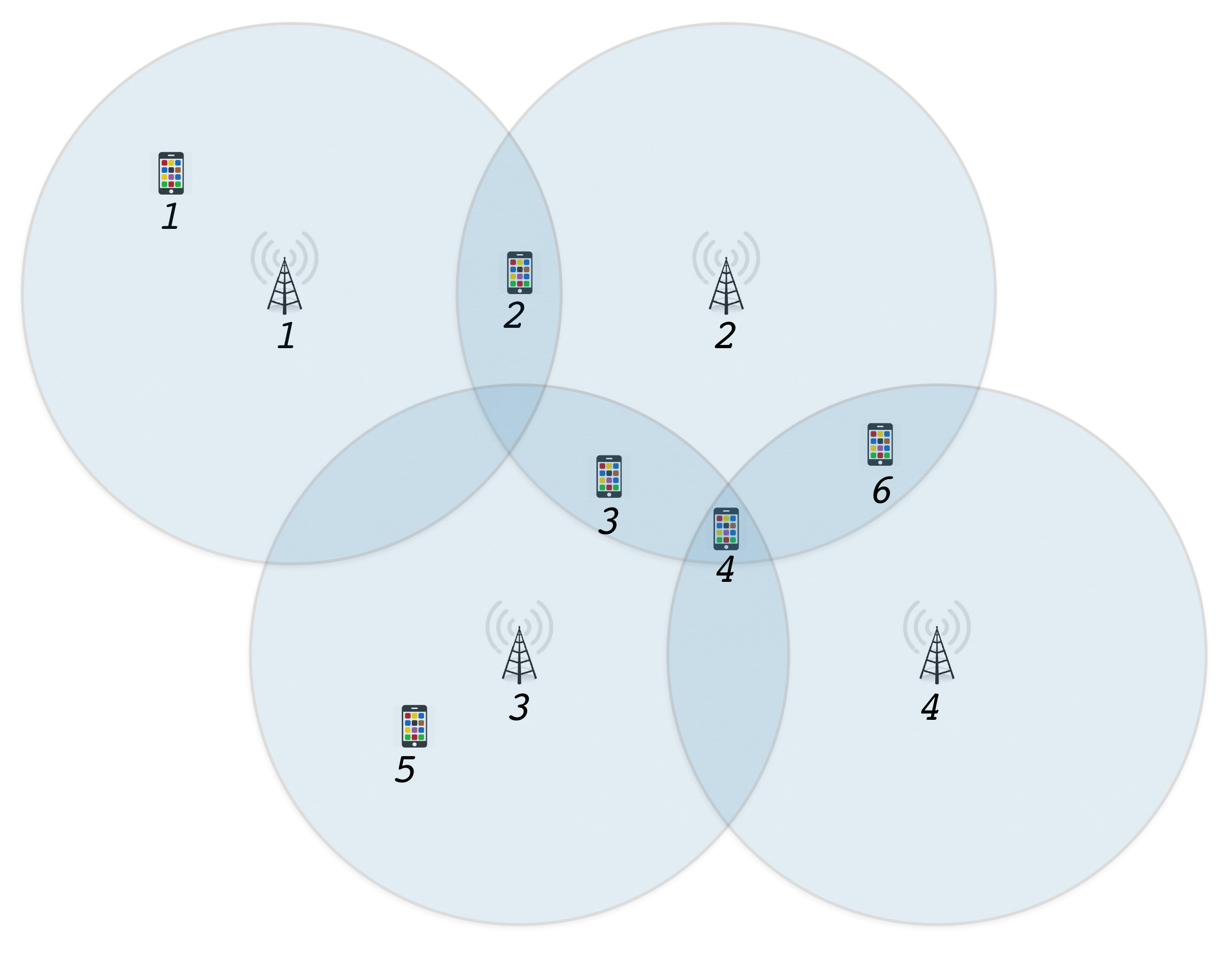}}  \hspace{2cm}
    \subfloat[Corresponding topological network graph.]{\includegraphics[width=7cm]{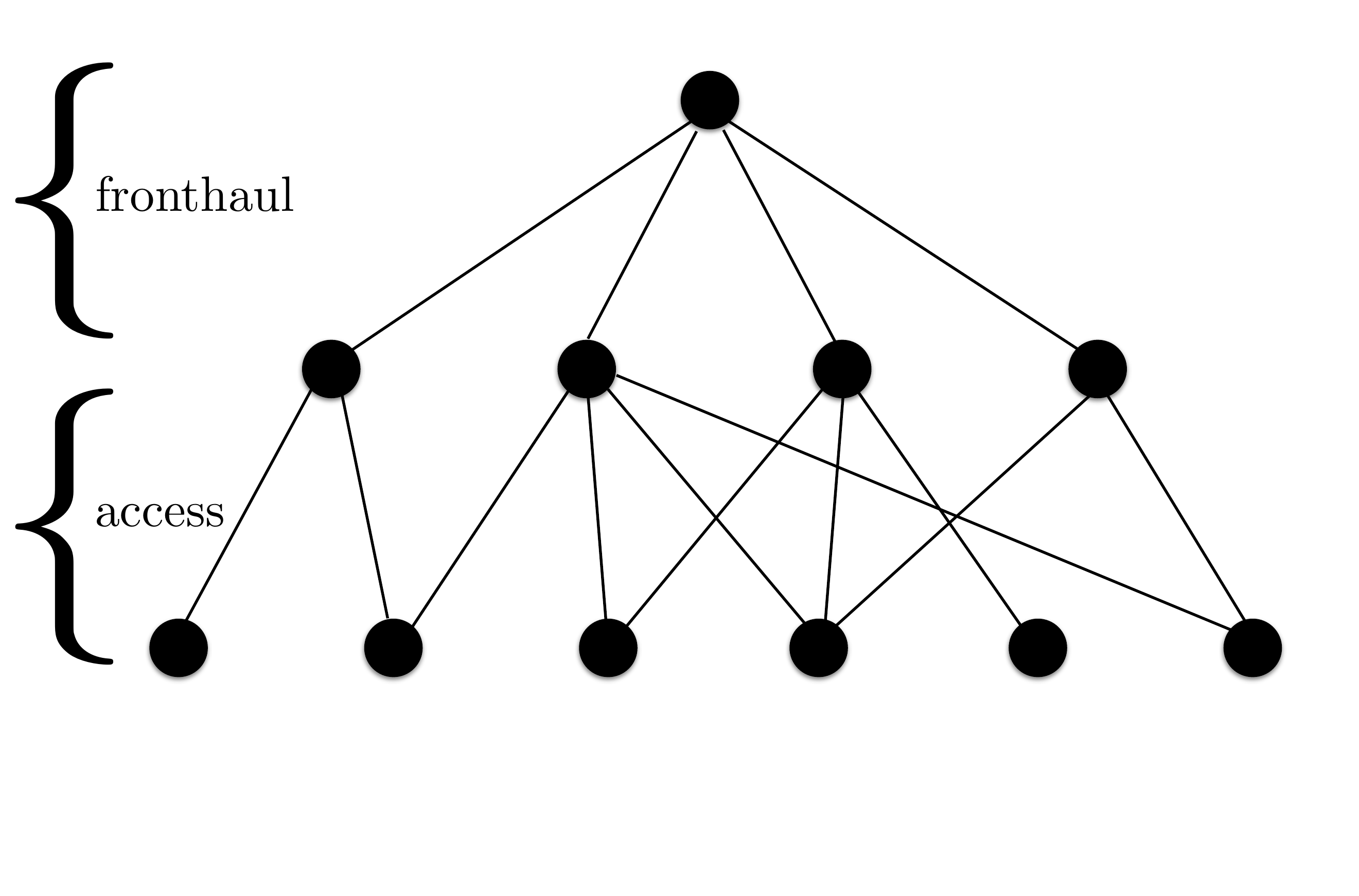}}
    \caption{An example of the network topology considered in Section \ref{sub:system} with $H=4$  and $K=6$. In the graph the helpers and the users are enumerated
    in increasing order from left to right.}
 \label{fig:scenario}
\end{figure}

In this section we consider our first network abstraction, formed by non-interfering links connecting the server to the helpers and the helpers to the users.
The network is defined by layered graph as shown in Fig.~\ref{fig:scenario}. The underlying topology of the spatially distributed wireless network is reflected in 
the association between users and helpers. In particular,  each helper is connected to a number of users 
through some mechanism of user-BS association (e.g., a helper is associated to all
the users within a certain {\em signaling radius} $a_{\rm sig}$).  
As motivated in Section \ref{sec:Introduction},  the specific association mechanism may be the result of 
some legacy PHY/MAC schemes  operating ``below IP'' and not under the control of the caching/content delivery system.  
For the sake of our treatment, the details of the user-helper association mechanism are irrelevant since 
the access network topology is not part of the delivery optimization. 
The links in this model are ``logical'', i.e., we assume that some reuse, resource allocation, and MAC protocol
are able to maintain given transmission rates between associated pairs of helpers and users, such that the network layer 
can execute a routing algorithm on the resulting graph. 
%Therefore, in the perspective of the routing algorithm, these links are {\em orthogonal}, i.e., 
%are not constrained by mutual interference (same output) or broadcast (same input) constraints.  
%This means that the links outgoing from the same helper can be driven by independent input data, and the links ingoing into the same user
%provide separate output data. 
The fact that users can receive simultaneously from multiple helpers reflects some form of {\em macro-diversity} and {\em carrier aggregation}, 
for which the users are able to receive on different channels at the same time.  
%At this abstraction level, it is irrelevant how the link orthogonality in the access segment 
%is achieved (for example, this may be the result of some TDMA, FDMA, CDMA scheme, 
%with some appropriate frequency reuse across different helpers). 

\subsection{System Model}
\label{sub:system}

We denote the library of $N$ files as $\mathcal{F} = \{W_1, W_2, \dots, W_N\}$ where each file has size of $F$ bits. 
The connectivity between the server and the helpers is referred to as the {\em fronthaul}, and consists of $H$ error-free wired (i.e., non-interfering) links of capacity
$C_{\rm front}$ bits per unit time  connecting the server to the helpers. 
The connectivity between helpers and users is referred to as the {\em access network} and consists of a bipartite graph (see Fig.~\ref{fig:scenario}). 
We assume that sum of the capacities of the links outgoing from any given helper cannot be larger than the downlink sum capacity of the helper. 
For simplicity, we consider the symmetric case where all helpers have the same downlink capacity denoted by $C_{\rm access}$.
This assumption can be trivially generalized to helpers with different downlink capacities.

In passing, we notice that the network model considered in this section is a generalization of the model in \cite{mejournal, mital2017coded}, where each user is connected to exactly  $r$ helpers (for some integer $r \in [H]$). In turns, this  is a generalization of the so-called {\em combination network}, 
where $K = {H \choose r}$ and there is a user connected to each distinct combination of $r$ out of $H$ helpers, 
as considered in the context of coded caching in  \cite{ji2015cachingasilomar,wan2017caching}.%,zewail2017coded,wan2017novel}.  
The subset of users connected to helper $h \in [H]$ and the subset of helpers connected to user $k \in [K]$  are denoted 
by $\mathcal{U}_h$  and $\mathcal{H}_k$, respectively.  
%%%%%%%%%%%%%%%%%%%%%%%%%%%%%%%%%%%%%%%%%%%%%%  
The capacities of the links $h \to k$, denoted by $C_{h \to k}$, must satisfy the sum constraint 
\begin{align} 
\sum_{k \in \mathcal{U}_h} C_{h \to k} \leq C_{\rm access}.  \label{sum-constraint}
\end{align}
Each user is equipped with a cache memory capable of storing up to $MF$ bits, for some $M \in [N]$, 
while the helpers do not possess any cache memory and can only forward packets received from the server to their connected users. 
In the prefetching phase, user $k\in[K]$ stores some subfiles from the $N$ library files.  
This phase is done without knowledge of the users' demands and of the network topology. 
We denote the content in the cache of user $k\in[K]$ by $Z_{k}$ and let $\Zm= \{Z_{1},\ldots,Z_{K}\}$ denote the collection of all caches in the system.
During the delivery phase, each user $k\in[K]$ demands file $W_{d_{k}}$ where $d_k \in[N]$. The demand vector $\dv= \{d_{1},\ldots,d_{K}\}$ is revealed to 
all nodes as metacontent information embedded in the packets sent by the server to the users.  
% delivery
Given $(\dv,\Zm)$, the server sends message $X_{s \to h}$ 
of $R_{h}F$ bits to helper $h$, for all $h\in [H]$. 
Then, helper $h$ transmits  message $X_{h\to k}$ 
of $R_{h\to k}F$ bits to user $k$, for all $k \in \Uc_h$. 
% decoding
User $k\in[K]$ must recover its desired file $W_{d_{k}}$ from $Z_{k}$ and the collection of received messages $\{X_{h\to k} : h\in \Hc_k\}$ from the helpers. 
A coded caching scheme is said to be {\em feasible} if, for any demand vector $\dv$, 
all users recover their desired file with vanishing probability of error as $F \rightarrow \infty$. 

As in  most literature on coded caching (e.g., see \cite{maddah2014fundamental,yu2017exact}),  
we focus on the worst-case delivery time over all possible demand vectors $\dv$. 
We consider a pipelined transmission in which fronthaul links and local access
links work in parallel. Messages are sufficiently long such that they can be broken into smaller packets and helpers can simultaneously
receive such packets from their fronthaul links while transmitting previously received packets on the access links to the users. 
When the number of packets per message is large, the delivery time is the maximum between the delivery times
in the first (fronthaul) and second (access) hops.   The latency along a path from the server to user $k \in \Uc_h$ via helper $h$ 
is given by  $\max \left \{ \frac{R_h F}{C_{\rm front}}, \frac{R_{h \rightarrow k} F}{C_{h \rightarrow k}} \right \}$.
For each user $k$, the overall delivery time is the maximum of the latencies incurred by the data that have to reach user $k$, given by 
$\max_{h \in \Hc_k}  \max \left \{ \frac{R_h F}{C_{\rm front}}, \frac{R_{h \rightarrow k} F}{C_{h \rightarrow k}} \right \}$. 
Finally, the worst-case user delivery time is given by 
\[ \max_{k \in [K]} \max_{h \in \Hc_k}  \max \left \{ \frac{R_h F}{C_{\rm front}}, \frac{R_{h \rightarrow k} F}{C_{h \rightarrow k}} \right \}, \]
which is equivalently written as $\max\{ T_{\rm front} , T_{\rm access}\}$, where we define
\[ T_{\rm front} = \max_{h\in [H]} \frac{ R_h F }{C_{\rm front}} \] 
and
\[ T_{\rm access} =  \max_{h\in [H], k\in \Uc_h} \frac{R_{h\to k} F}{ C_{h \to k}}. \]
Eventually, the minimization over the cache design $\Zm$ and the access link capacity allocation $\{C_{h \rightarrow}\}$ of the worst-case user delivery time is given by 
\begin{align}%{rCl}
T^{\star}
=
\min_{\substack{\Zm}}
\max_{ \dv\in[N]^{K}} \min_{\substack{\{C_{h \to k}\} \\ \sum_{k } C_{h \to k} \leq C_{\rm access}}} 
\max\left\{ T_{\rm  front}, T_{\rm access} \right \}.
\label{fuckme}
\end{align}

%%%%%%%%%%%%%%%%%%%%%%%%%%%%%%%%%%%%%%%%%%%%%%%%%
\subsection{MAN Caching Scheme}
\label{sub:MAN}

For future reference, we briefly review here the MAN scheme for single shared-link network \cite{maddah2014fundamental}.
This network is a special case of the topological network model treated here with $H = 1$ helper connected to all $K$ users, 
and with $C_{\rm access} = K C_{\rm front}$.  Define the {\em library replication parameter} $t = KM/N$ as how many times the library can be contained in the collective cache memory of all users,  and assume that $t \in [0:K]$ (for non-integer $t$, a standard memory sharing approach can be used as done in several existing works, e.g.,  \cite{maddah2014fundamental}).  %}{\RED [put ref]}).  
 We define the collection of all user subsets of some integer size $r \in [0:K]$ as $\Omega^K_r = \{ \Uc \subseteq [K] : |\Uc | = r\}$. 
Each file $W_i$ is partitioned into $\binom{K}{t}$ non-overlapping and equal-length subfiles $W_{i,\Tc}$, for all user subsets $\Tc \in \Omega^K_t$. 
Each user $k\in [K]$ caches the subfiles $W_{i, \Tc}$ for all  $\Tc \ni k$ for all $i \in [N]$.  
In the delivery phase, for each subset $\Sc \in \Omega^K_{t+1}$, 
the server broadcasts the coded multicast message 
\begin{equation}
\label{eq:coded_multicast}
	V_{\Sc} = \bigoplus_{k \in \Sc} W_{d_k, \Sc\setminus\{k\}}.
\end{equation} 
Each user $k\in \Sc$ requires $W_{d_k, \Sc\setminus\{k\}}$ since by definition this subfile is needed ($d_k$ is the index of the file wanted by user $k$) and not
 cached (since obviously $k \notin \Sc \setminus \{k\}$). Furthermore, all other subfiles  $W_{d_j, \Sc\setminus\{j\}}$ for $j \in \Sc, j \neq k$ 
in the XOR (\ref{eq:coded_multicast}) 
are in the cache of user $k$ since $k \in \Sc\setminus\{j\}$. Therefore, each user $k \in \Sc$ can recover
$W_{d_k, \Sc\setminus\{k\}}$ from $V_{\Sc}$ and eventually all demands are satisfied. 
For sufficiently large $F$, so that the subpacketization order ${K \choose t}$ is possible, then the MAN scheme achieves deliver time %[REF MAN]
\begin{equation} 
T^{\rm MAN} = \frac{F}{C_{\rm front}} \frac{{K \choose t+1}}{{K \choose t}} = \frac{F}{C_{\rm front}} \frac{K - t}{1 + t}. 
\end{equation}

%%%%%%%%%%%%%%%%%%%%%%%%%%%%%%%%%%%%%%%%%%%%%%%%%%%% 
\subsection{Centralized Coded Caching and Routing Optimization}
\label{sec:MainResult}

As a prelude to the proposed caching and delivery scheme of Section \ref{sec:grouping}, we present here a direct 
application of the MAN scheme with routing-based delivery for a general topological network model. 
The purpose of this section is also to illustrate why such scheme would be completely impractical, and therefore to motivate the following 
novel schemes.   Define $t$ and the caches configurations $\{Z_k\}$ as in the MAN scheme of Section \ref{sub:MAN}. 
%, assume $t \in [0:K]$, consider the subfiles $\{W_{i, \Tc} : \Tc \in \Omega^K_t, i \in [N]\}$ and let
%each user cache $Z_k = \{ W_{i, \Tc} :   \Tc \in \Omega^K_t, \Tc \ni k, i \in [N] \}$.
At the server side, the subfiles are individually precoded by using an erasure code over a sufficiently large finite field. 
Each subfile is represented as a sequence of finite field symbols over a binary-extension 
finite field $\FF_{2^q}$ for some integer $q$. We denote by $\widehat{W}_{i,\Tc}$ the encoded version of $W_{i,\Tc}$, i.e., 
the resulting sequence of linear combinations over $\FF_{2^q}$.~\footnote{ In this section with use the ``hat'' notation $\widehat{A}$ to indicate blocks $A$ of symbols over $\FF_q$, whose length $|\widehat{A}|$ is given by the number of 
 finite-field symbols.}
 Letting $F' = F/(q{K \choose t})$ denote the length of the subfiles $W_{i,\Tc}$ in finite-field symbols, we define the normalized 
 length of the codewords $\widehat{W}_{i,\Tc}$ as $y = |\widehat{W}_{i,\Tc}|/F'  \geq 1$, same for all
subfiles.  This precoding is designed such that the original subfile $W_{i,\Tc}$ can be retrieved from any $F'$ distinct 
symbols of $\widehat{W}_{i,\Tc}$.\footnote{This can be implemented by 
intra-session random linear network coding~\cite{randomnetworkcoding,wan2017novel,Mital2017randomtopo}
or algebraic MDS codes, as long as an MDS code with parameters $(F', y F')$ over $\FF_{2^q}$ exists.} 

In the delivery phase, we create the coded multicast messages for each user group $\Sc \in \Omega^K_{t+1}$ 
by XOR-ing the precoded subfiles, i.e., we let $\widehat{X}_{\Sc} =  \bigoplus_{k \in \Sc} \widehat{W}_{d_k, \Sc\setminus\{k\}}$.
We let $\widehat{X}_{\Sc}^h$ denote the segment of $\widehat{X}_{\Sc}$ sent to helper $h$. 
After receiving $\widehat{X}^h_{\Sc}$, helper $h$ forwards it to the users in $\Sc \cap \Uc_h$.  
We let $y_{\Sc}^h  = \frac{|\widehat{X}^h_{\Sc}|}{F'}$,  where $y_{\Sc}^h=0$ if $\Sc \cap \Uc_h = \emptyset$, 
 i.e., $\widehat{X}^h_{\Sc}$ is not forwarded at all to the helpers not connected to at least one user in $\Sc$.
When a user $k \in \Sc$ receives  $\widehat{X}^h_{\Sc}$ containing some symbols 
of its desired codeword $\widehat{W}_{d_k, \Sc\setminus\{k\}}$, it is able to locally generate the corresponding 
symbols of the interfering codewords $\widehat{W}_{d_j, \Sc\setminus\{j\}}$ for $j \in \Sc, j \neq k$ participating in the XOR form its cache content, 
and ``cache out'' the desired symbols. Thanks to the subfile precoding, if a user recovers $F'$ distinct symbols of 
$\widehat{W}_{d_k, \Sc\setminus\{k\}}$, then it will be able to decode the whole desired subfile $W_{d_k, \Sc\setminus\{k\}}$. 
By making $y$ large enough, it is always possible to make sure that the symbols cached out by user $k$ from all its connected helpers $h \in \Hc_k$
are all distinct. It follows that the scheme is feasible if
\begin{equation}
\label{eq:cons}
\sum_{h \in \mathcal{H}_k}  y_{\Sc}^h \geq 1,  \quad \forall \; \Sc \in \Omega^K_{t+1}, \;\; \mbox{and} \; \forall \; k \in \Sc. 
\end{equation}
From the above arguments it is clear that the scheme is feasible for sufficiently large field size $2^q$ and sufficiently large normalized 
length $y$ since, for any $\dv$, each user $k\in [K]$ can decode its desired file $W_{d_k}$  if 
the server sends messages $X_\Sc^h$ to helpers $h \in [H]$ for all $\Sc \in \Omega^K_{t+1}$ such that  \eqref{eq:cons} holds.

%NOTICE
 
The length (in bits) of the message $\widehat{X}^h_{\Sc}$ is given by $q F' y^h_{\Sc}  =   F y^h_{\Sc} /{K \choose t}$. 
The normalized link load from the server to each helper $h\in [H]$ is obtained by summing the length of all 
the multicast messages and dividing by $F$, yielding
\begin{equation}
R_h = \frac{1}{{K \choose t}} \sum_{\Sc \in \Omega^K_{t+1}} y^h_{\Sc}. 
\end{equation}
Similarly, the normalized load for link $h \to k$ is obtained by summing over the messages forwarded by $h$ to $k$ and is given by 
\begin{equation}
R_{h\to k} = \frac{1}{{K \choose t}} \sum_{\Sc \in \Omega^K_{t+1} :  \Sc \ni k} y^h_{\Sc}. 
\end{equation} 
Hence, we have
\begin{align}
T_{\rm front}  & = \max_{h \in [H]}  \frac{F}{C_{\rm front} {K \choose t}}  \sum_{\Sc \in \Omega^K_{t+1}} y^h_{\Sc},  \label{eq:time_f}   \\
T_{\rm access}  & = \max_{h\in[H]} \max_{k\in\mathcal{U}_h}  \frac{F}{C_{h \to k} {K \choose t}}  \sum_{\Sc \in \Omega^K_{t+1} :  \Sc \ni k} y_{\Sc}^h,   \label{eq:time_e}
\end{align}
and resulting routing and resource allocation problem is given by 
\begin{subequations}
\label{eq:opt_time}
\begin{align}
~ \underset{(y_{\Sc}^h, C_{h \to k} \geq 0 \; : \; \Sc \in  \Omega^K_{t+1} , h \in [H], k \in[K] )}{\text{minimize}} &   \max \{T_{\rm access} ,T_{\rm front}\}\\
 \quad \quad \quad \quad  \text{subject to:} \quad 
& y_{\Sc}^h=0,   ~ \forall ~ \Sc \cap \Uc_h = \emptyset, \\ 
&  \sum_{h \in \mathcal{H}_k}^{} y_{\Sc}^h \geq 1 , ~  \forall \; \Sc \; \mbox{and} \; \forall \; k \in \Sc, \\
& \sum_{k \in \mathcal{U}_h}^{} C_{h \to k} \leq C_{\rm access}, \forall h \in [H].
\end{align}
\end{subequations}
The optimization in \eqref{eq:opt_time} is not an LP because the objective function contains the ratio $\frac{y_{\Sc}^h}{C_{h \to k}}$. 
However, it is possible to solve (\ref{eq:opt_time}) through a sequence of LPs.  First, we rescale the objective function in (\ref{eq:opt_time}) 
by $C_{\rm front} {K \choose t} /F$, which is just a fixed constant. Up to this scaling,  the problem is equivalent to
 \begin{subequations}
 \label{eq:opt_dummy}
 \begin{align}
  \underset{(y_{\Sc}^h, C_{h \to k} \geq 0 \; : \; \Sc\in  \Omega^K_{t+1} , h\in [H], k\in[K] )}{\text{minimize}}   \alpha  &\\
   \quad \quad \quad \quad \text{subject to:}   
&   \sum_{\Sc \in \Omega^K_{t+1}} y_{\Sc}^h \leq \alpha,  ~\forall h \in [H],  \\
&  \sum_{\Sc \in \Omega^K_{t+1}:  \Sc \ni k} y_{\Sc}^h \leq \frac{C_{h \to k}}{C_{\rm front}} \alpha, ~\forall h \in [H], \forall k \in [K],  \\
&  y_{\Sc}^h=0,   ~ \text{if } \Sc \cap \Uc_h = \emptyset, \\ 
&   \sum_{h \in \mathcal{H}_k}^{} y_{\Sc}^h \geq 1 , ~  \forall \; \Sc \; \mbox{and} \; \forall \; k \in \Sc, \\
&    \sum_{k \in \mathcal{U}_h}^{} C_{h \to k} \leq C_{\rm access}, \forall h \in [H].
 \end{align}
 \end{subequations}
 % delivery time = alpha /(C_F/F')

Written in the form (\ref{eq:opt_dummy}), the problem can be solved by 
considering an interval $\alpha \in [0, \bar{\alpha}]$ large enough such that the feasibility problem associated with 
the constraints in \eqref{eq:opt_dummy} for fixed $\alpha = \bar{\alpha}$ is satisfied.  Then, using the bisection method,  
we can determine the minimum $\alpha$ for which feasibility is satisfied.~\footnote{In practice, the search stops when the gap between minimum feasible $\alpha$ and the maximum unfeasible $\alpha$ is small enough.} 

As previously anticipated, the subpacketization order of the MAN scheme becomes quickly very large since
$|\Omega^K_{t}  | \geq 2^{K h_2(M/N)}$ (where $h_2(\cdot)$ is the binary entropy function). 
For constant fractional cache memory $M/N = \mu$ and a large number of users, 
this exponential growth in $K$ makes the scheme impractical. 
For example, for a system with $K = 1000$ users and $\mu = 0.01$ (each user caches 1\% of the library), 
the number of subpackets is larger than $2 \times 10^{24}$. Even computing the solution of problem (\ref{eq:opt_dummy}) becomes intractable, 
because the number of variables $\{y_{\Sc}^h\}$ and constrains is larger than the number of XOR messages $|\Omega^K_{t+1}|$.  
In the next section we address both these problems. 
%In the new scheme proposed in the next section, we address  both the subpacketization order and the decentralized prefetching problems. 
%As a desirable byproduct, we also obtain routing optimization problems with linear complexity in the number of users $K$.

%%%%%%%%%%%%%%%%%%%%%%%%%%%%%%%%%%%%%%%%%%%%%%%%%%%%%%%
\subsection{Decentralized Coded Caching with Cache Replication}
\label{sec:grouping}

In order to reduce the subpacketization order and allow for {\em decentralized} prefetching, such that users can join 
and leave the system at any time irrespectively of the other users,   we consider the cache replication approach of \cite{jin2016} in the context 
of our ``network layer'' model and propose two delivery schemes.  
The first is a direct extension of the so-called {\em multiround} delivery scheme of \cite{parrinello2018fundamental}. The second is the first important 
novel contribution of this paper, and turns out to be generally more efficient.
Following \cite{jin2016}, we fix an integer $L < K$ and create a MAN subpacketization for the a system with $L$ {\em virtual} users, with 
library replication parameter $t' = LM/N \in [0:L]$. 
Hence, each file $W_i$ is divided into $\binom{L}{t'}$ non-overlapping  and equal-length subfiles $\{W_{i,\Tc}: \Tc \in \Omega^L_{t'} \}$. 
We generate $L$ cache configurations,  one for each virtual user, according to the MAN scheme for $L$ users, 
such that the $\ell$-th cache configuration is given by $\widetilde{Z}_\ell = \{ W_{i, \Tc} :  \Tc \in \Omega^L_{t'}, \Tc \ni \ell, i \in [N] \}$. 
Notice that the subpacketization reduces from $\binom{K}{t}$ to $\binom{L}{t'}$. 
For example, in a system with $K = 1000$ users and $\mu = 0.01$ and $L = 10$ the subpacketization order reduces 
to $\sim 300$ instead of $\sim 2 \times 10^{24}$. 

In the decentralized prefetching phase each user $k$, when joining the system and independently of the other users, picks at random an index $\ell \in [L]$ and loads the cache configuration $\widetilde{Z}_{\ell}$, i.e., it lets $Z_k = \widetilde{Z}_{\ell}$. 
The set of users with cache configuration  $\ell$  is denoted by $\Pc_{\ell}$. By construction, all users $k \in \Pc_{\ell}$ have the same cache configuration 
$Z_k = \widetilde{Z}_{\ell}$, and the user groups define the partition $\Pc = \{\Pc_1, \Pc_2, \dots, \Pc_L\}$.
In the following, we describe two delivery schemes for the system with cache replication and decentralized prefetching. 
%The first scheme is  based on the   multiround delivery idea of \cite{parrinello2018fundamental, jin2016} applied to our network, 
%where for each round an optimization problem of the type (\ref{eq:opt_dummy}) can be solved over a much smaller system with at most $L$ users. 
%The second scheme is entirely novel, and further improves upon the first. 

%%%%%%%%%%%%%%%%%%%%%%%%%%%%%%%%%%%%%%%%%%%%%%%%%%%%%%%%%%
\subsubsection{Multiround Delivery with Routing Optimization}
\label{sub:direct extension}

We start by recalling the so-called {\em multiround delivery} of \cite{parrinello2018fundamental, jin2016} to handle the case of cache replication for the 
MAN model (shared-link network). Consider the case where a single server is connected to $K$ users partitioned into
$L$ caching groups $\Pc_1, \Pc_2, \dots, \Pc_L$ as said before, and define the sorting permutation $[\cdot]$ such that
$|\Pc_{[1]}| \geq |\Pc_{[2]}| \dots \geq |\Pc_{[L]}|$. Define the {\em delivery array}
$\Bm$ with  $L$ rows and $|\Pc_{[1]}|$ columns formed by placing in each row $\Bm_{\ell, :}$ the users in $\Pc_{[\ell]}$ for $\ell = 1, \ldots, L$, respectively, 
where rows $\ell$ with $|\Pc_{[\ell]}| < |\Pc_{[1]}|$ are padded by zeros in order to have all rows of the same length. 
The ordered numbers $|\Pc_{[1]}| \geq |\Pc_{[2]}| \dots \geq |\Pc_{[L]}|$, i.e., the length of the non-zero leading segment of each row of $\Bm$, are referred to
as the {\em occupancy numbers} of the delivery array. 
Then, each column $\Bm_{:,j}$, for $j = 1, \ldots, |\Pc_{[1]}|$ corresponds to a set of users with distinct cache configurations. 
If $\Bm_{\ell,j} = 0$ it means that the user with the $\ell$-th cache configuration is not present. 
For each column $j$, let $\Rc_j = \{ \ell \in [L] : \Bm_{\ell, j} \neq 0\}$ denote the set of present cache configurations. 
The multiround delivery serves the users in each column of the array $\Bm$  (i.e., each delivery round) 
by forming the XORs corresponding to the MAN scheme with parameters $N$ files, $L$ users and library replication $t'$, 
only for the multicast subsets $\Sc \in \Omega^L_{t'+1}$  such that $\Sc \cap \Rc_j \neq \emptyset$ (otherwise the XOR messages would be useless 
in round $j$). Each XOR packet of such delivery has length $F/{L \choose t'}$ bits, and all the users in column $j$ are served
in ${L \choose t'+1} - {L - |\Rc_j| ] \choose t'+1}$ XOR transmissions. 
Grouping the columns by the number of zeros $b = 0,\ldots, L-1$, 
$\Bm$ contains exactly $|\Pc_{[L-b]}| - |\Pc_{[L-b+1]}|$ columns with $b$ zeros (where we define $|\Pc_{[L+1]}| := 0$). 
The duration of the multiround delivery in XOR packets is
\begin{equation}
\sum_{b=0}^{L-1} |\Pc_{[L-b]}| - |\Pc_{[L-b+1]}| \left ( {L \choose t'+1} - {b \choose t'+1} \right ).
\end{equation}
Dividing by ${L \choose t'}$ and rearranging terms, we arrive at the total delivery load 
 \begin{equation}
     \label{eq:eliarate}
     R_{\rm multiround} = \sum_{r = 1}^{L-t^{\prime}} \frac{|\Pc_{[r]}| \binom{L-r}{t^{\prime}}}{\binom{L}{t^{\prime}}}.
 \end{equation}
The load in (\ref{eq:eliarate}) was proven to be optimal for the shared-link network with assigned occupancy numbers \cite{parrinello2018fundamental}.

In order to extend multiround delivery to the considered two-hop helper network, after partitioning the users into the caching groups
$\Pc_1, \Pc_2, \dots, \Pc_L$, we can apply the routing optimization (\ref{eq:opt_dummy}) for each delivery round 
(i.e., column of the resulting delivery array $\Bm$) independently,  with the caveat that the set of XORs in round $j$ is given by 
$\{\Sc \in \Omega^L_{t'+1} : \Sc \cap \Rc_j  \neq \emptyset \}$.
Eventually,  the multiround scheme yields a total of $|\Pc_{[1]}|$  separate  optimization problems, i.e., one for each round.
Notice that $|\Pc_{[1]}| \leq K$ (in fact, for large $K$ and random uniform assignment of the caching groups this is close to $K/L$ 
up to small fluctuations by the law of large numbers) and  $|\Omega^L_{t'+1}| = O(1)$ (constant with respect to $K$). Therefore, 
the complexity of this scheme is {\em linear} in the number of users $K$. 

%%%%%%%%%%%%%%%%%%%%%%%%%%%%%%%%%%%%%%%%%%%%%%%%%%%%%%%%%%%%%%%%
\subsubsection{New Delivery Scheme Based on Routing Optimization}
\label{sub:novel decent method}

Although multiround delivery is worst-case load optimal for the single shared-link network, 
in our model users with the same cache configuration are not equivalent due to the network topology.  
Hence, the load depends on the choice of the users in each round.  
%Finding the best users-to-rounds assignment is an integer program of prohibitive complexity in practical scenarios.
%In addition, even if we could find the best assignment, the delivery organized in multiple rounds of MAN scheme 
%may be still suboptimal over all possible delivery schemes. 
Therefore, applying multiround delivery for our topological network model
is generally suboptimal. 

In this section we propose a novel delivery strategy that generally beats multiround delivery. 
The idea is to create coded multicast messages based on the network topology, such that they are simultaneously useful for users connected to same helper.  
In other words, there are no multicast messages simultaneously useful for users $k,k'$ for which $\Hc_k \cap \Hc_{k'} = \emptyset$ 
(no common helpers in their connectivity set).  For each user $k$ in caching group $\ell_k$, we need to transmit missing subfiles of the type 
$W_{d_k,\Sc \setminus \{\ell_k \}}$  where $\Sc \in \Omega^L_{t'+1} \text{ and }  \Sc \ni \ell_k $.
We define $ {W}_{d_k,\Sc \setminus\{\ell \}}^h$  as the segment of ${W}_{d_k,\Sc \setminus\{\ell \}}$ sent to helper $h \in \Hc_k$ and define
$y^h_{k, \Sc \setminus \{\ell \} }  = \frac{|{W}_{d_k,\Sc \setminus\{\ell \}}^h|}{{ F}/\binom{L}{t'}}$ as its normalized length. 
We define $\widetilde{W}_{\ell, \Sc \setminus\{\ell\}}^h = \cup_{k \in \Pc_\ell \cap \Uc_h}  {W}^h_{d_k,\Sc \setminus\{\ell\}}$ 
as the concatenation of the subfiles segments required by the users connected to helper $h$ with same cache configuration $\ell$. 
Notice that if $\Pc_\ell \cap \Uc_h = \emptyset$ then $\widetilde{W}_{\ell, \Sc \setminus\{\ell\}}^h$ is empty and therefore has zero length. 
The coded multicast message to helper $h$ is obtained as follows
\begin{equation}
X_{\Sc}^h = \underset{\ell  \in \Sc}{\oplus } \widetilde{W}_{\ell, \Sc\setminus \{\ell\}}^h.    \label{sucamillo}
\end{equation}
 Since the sets $\Pc_\ell \cap \Uc_h$ may have different sizes, the concatenated segments 
$\widetilde{W}_{\ell, \Sc\setminus \{\ell\}}^h$ may have different lengths. Then, zero-padding is used in (\ref{sucamillo}) such that 
$X_{\Sc}^h$ has length 
\begin{equation} 
|X_{\Sc}^h| = \max_{\ell \in \Sc} |\widetilde{W}_{\ell, \Sc\setminus \{\ell\}}^h| =   \max_{\ell \in \Sc} \sum_{k \in \Pc_\ell \cap \Uc_h}  |{W}^h_{d_k,\Sc \setminus\{\ell\}}|.
\label{porcamadonna}
\end{equation}
For each $\Sc \in \Omega^L_{t'+1}$, helper $h$ forwards to user $k \in \Uc_h \text{ and } \Sc \ni \ell_k$ just the useful portion of $X^h_\Sc$, 
of length $|{W}^h_{d_k,\Sc \setminus\{\ell_k\}}|$. 
In any case, the bits of ${W}^h_{d_k,\Sc \setminus\{\ell_k\}}$  can be cached out from this portion since all other interfering bits come from subfiles 
present in cache $Z_k = \widetilde{Z}_{\ell_k}$.

Consistently with the definitions introduced before and using (\ref{porcamadonna}), we have 
 \begin{align}
 & T_{\rm front} = \max_{h\in [H]}  \frac{F}{C_{\rm  front}  \binom{L}{t'} } \sum_{\Sc \in \Omega_{t'+1}^L}  \max_{\ell \in \Sc}  \sum_{k \in \Pc_\ell \cap \Uc_h}  y^h_{k, \Sc \setminus \{\ell\}} \label{eq:T front1} \\
  & T_{\rm access } = \max_{h \in [H]} \max_{k \in \Uc_h} \frac{F}{C_{h \to k} \binom{L}{t'}} 
  \sum_{\Sc \ni \ell_k} y^h_{k, \Sc \setminus \{\ell_k\} } 
 \label{eq:T access1}
 \end{align}
 The resulting routing optimization problem is given by 
\begin{subequations}
 \label{eq:opt_dummy1}
 \begin{align}
 ~ \underset{(y_{k,  \Sc \setminus \{\ell_k\} }^h,  C_{h \to k} \geq 0): \Sc \in  \Omega_{t'+1}^L, h \in [H] , k \in [K])}{\text{minimize}}  
 	& \max \{T_{\rm front},T_{\rm access }\} \\
  \quad \quad \quad \quad \quad \text{subject to:} 
&  \sum_{h \in \Hc_k} y^h_{k, \Sc \setminus \{\ell_k\}}=1, \ { \forall k \in [K]}, \forall \Sc \ni \ell_k\\
&  \sum_{k \in \Uc_h}^{} C_{h \to k} \leq C_{\rm access}, \forall h \in [H].
 	\end{align}
 \end{subequations}
The problem reduces to a sequence of LPs in the same way seen for problem 
(\ref{eq:opt_dummy}) (details are omitted for the sake of brevity). 
Different from the delivery scheme in Section \ref{sec:MainResult}, where we use random linear network coding to encode subfiles, 
here this is not needed since the scheme is able to deliver requested subfiles in sequence of segments of original subfiles, avoiding any overlap.  
The number of variables in problem (\ref{eq:opt_dummy1}) is at most $K H \left( {L - 1 \choose t'} +  1\right)$ and the number of constraints 
at most $K {L - 1 \choose t'}  + H$. Therefore, also this problem has linear complexity in $K$.

%%%%%%%%%%%%%%%%%%%%%%%%%%%%%%%%%%%%%%%%%%%%%%%%%%%%%%%%%%% 
\section{Networks with Broadcast Constraints and Collision Interference}
\label{sec:broadcast-collision}

While in Section \ref{sec:Topological-networks} and \ref{sec:grouping} we assumed a model where transmission resources 
are already negotiated  according to some PHY/MAC protocol, such that the network layer ``sees'' the access 
network as a set of logical orthogonal links, in this section we consider a model which is one step closer to an actual 
physical wireless network, capturing the broadcast nature of the wireless medium and 
interference in the form of {\em collisions}.  

%%%%%%%%%%%%%%%%%%%%%%%%%%%%%%%%%%%%%%%%%%%%%%%%%%%%%%%%%%%%%%%%
%%%%%%%%%%%%%%%%%%%%%%%%%%%%%%%%%%%%%%%%%%%%%%%%%%%%%%%%%%%%%%%%
\subsection{System Model}
\label{sub:system1}

\begin{figure}
    \centering
    \subfloat[Layout geometry with broadcast transmission from the helpers and collision interference at the users.]{\includegraphics[width=6cm]{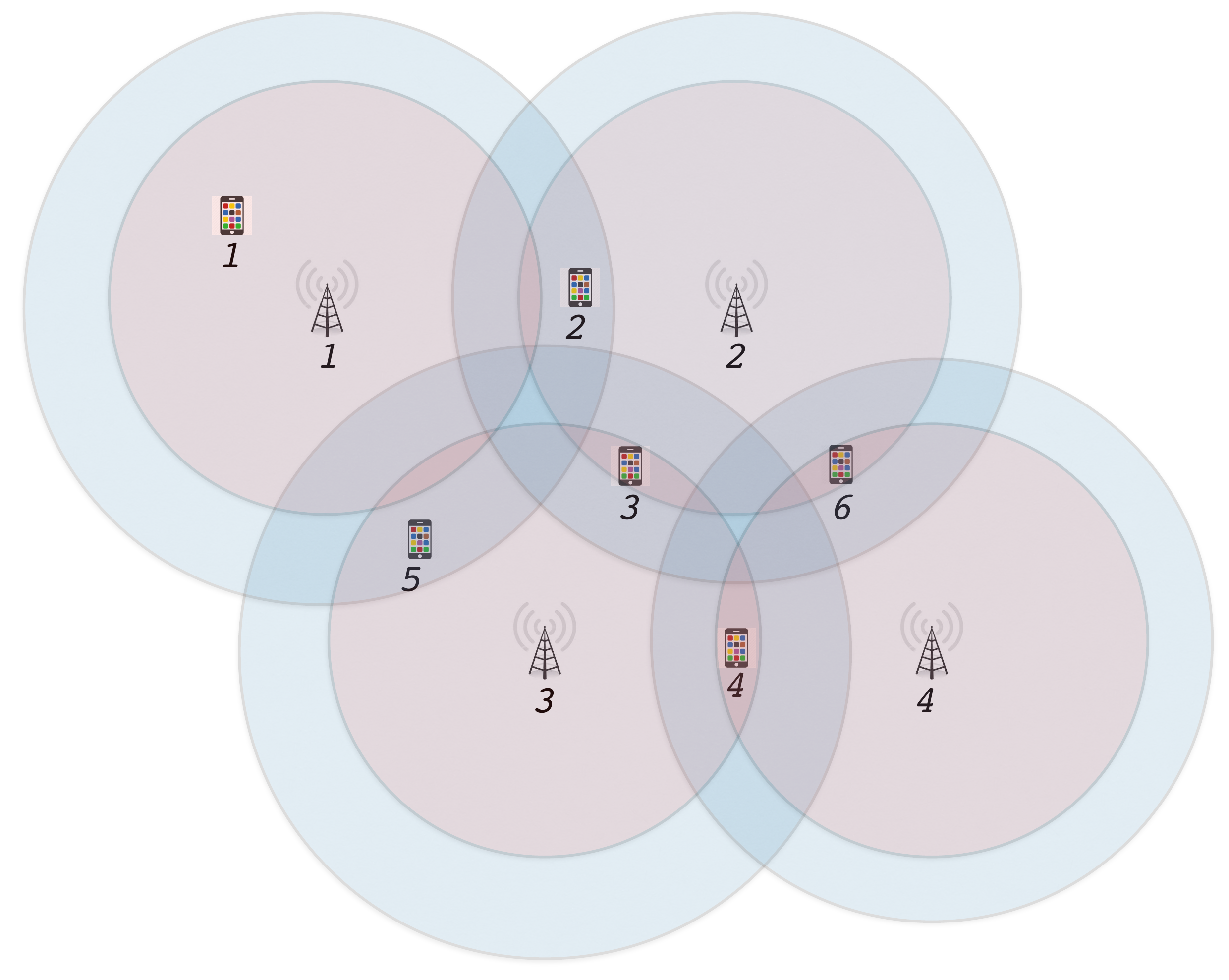}}  \hspace{2cm}
    \subfloat[Corresponding network graph with broadcast and interference constraints and A reuse scheme with $r = 3$ colors eliminating the interference conflicts.]{\includegraphics[width=7.5cm]{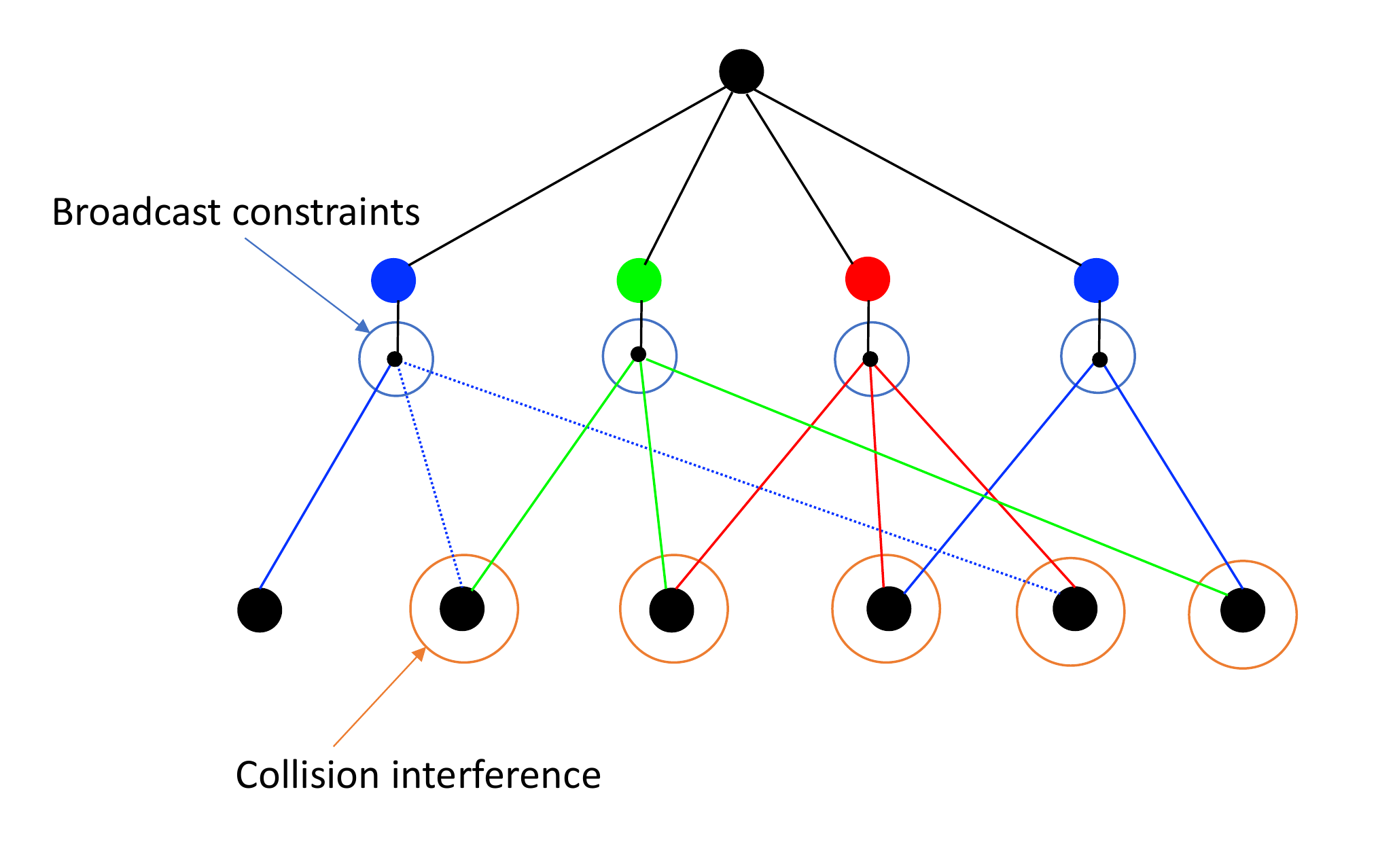}}
    \caption{An example of the network topology considered in Section \ref{sub:system} with $H=4$  and $K=6$. In the graph the helpers and the users are enumerated
    in increasing order from left to right. The dashed edges in the graph represent interference conflicts.}
 \label{fig:scenario1}
\end{figure}

Consider again the 2-dimensional plane geometry qualitatively illustrated in Fig.~\ref{fig:network-concept} and assume spatially distributed helpers and users. 
We refer to the helpers that actually transmit  as the ``active helpers'' in a given time slot.
The set of active helpers can be scheduled over different time slots according to some transmission strategy which is part of the 
system optimization studied in this section.  Since the radio channel is a broadcast medium, when a helper is active,
its transmission is received at all users within a certain radius $a_{\rm interf}$. 
Furthermore, we assume that users within a smaller  radius $a_{\rm cell} \leq a_{\rm interf}$ can successfully decode 
the helper message unless they are not interfered by some other (active) helper. 
%The shared medium property of the physical transmission is modeled as a {\em broadcast constraint} at the helper output: 
%namely, the symbols sent by  helper $h$ to all the users $k$ located within radius $a_{\rm cell}$ 
%are the same (there are no already granted ``orthogonal links'' $h \to k$).  
The decodability condition over the access network is formally expressed as follows: if user $k$ is located at distance not larger than $a_{\rm cell}$ from active helper $h$,  and there is no other active helper $h' \neq h$ within distance $a_{\rm interf}$ from user $k$, then user $k$ can decode the message of helper $h$. 
\footnote{This model has been widely used in the wireless networks literature and it is commonly referred as the {\em protocol model} 
(e.g., see \cite{gupta2000capacity}).} 
An example of the network  model considered in this section and the associated graph involving both interference conflicts 
and broadcast constraints is given in Fig.~\ref{fig:scenario1}.
Let $\Theta = \{\theta_h\} \subset \RR^2$ and $\Phi = \{\phi_k\} \subset \RR^2$ denote the set of helpers and users positions 
on the coverage area (a region in the 2-dimensional plane $\RR^2$), respectively. 
In order for the system to be feasible, it is necessary that any user location $\phi_k$ is covered by at least one disk 
of radius $a_{\rm cell}$ centered around a helper position $\theta_h \in \Theta$, i.e., letting
$\Bc(x, \rho)$ the disk centered at $x$ of radius $\rho$, a necessary condition for feasibility is that
\begin{equation} 
\phi_k \in \bigcup_{\theta_h \in \Theta_H} \Bc(\theta_h, a_{\rm cell}), \;\;\; \forall \; \phi_k \in \Phi.
\end{equation}
Since here we are not concerned with issues such as coverage probability or outage probability (widely studied in stochastic geometry \cite{haenggi2012stochastic }), here we make the assumption that this condition is satisfied. This corresponds to the fact that, realistically,  
the network layer takes  into consideration only the users that are actually associated to at least one helper and can thus receive data. 
As in the model of Section \ref{sec:Topological-networks}, we consider individual orthogonal fronthaul links of capacity $C_{\rm front}$  
connecting the server to the helpers, and let the downlink multicast rate of transmission from the (active) helpers be equal to $C_{\rm access}$. 
For simplicity, we use the same symbol $C_{\rm access}$ for the helpers downlink sum/multicast capacity.
%although these values need not be the same for the two models since they have different physical meanings. 
%More discussion on this issue is given in Section \ref{sec:NumericalResults}, where we discuss some connections between the two models. 
For the network model defined above, we consider a caching/delivery scheme where the prefetching is identical to the 
decentralized cache replication scheme of Section \ref{sec:grouping}  with $L$ caching groups, and use the same notation $t' = ML/N$ (integer value in $[L]$) 
and caching groups $\Pc_1, \ldots, \Pc_L$ introduced in  Section \ref{sec:grouping}. 
The proposed delivery schemes for the centrally computed coded caching multicast messages is discussed in the following subsections. 

%%%%%%%%%%%%%%%%%%%%%%%%%%%%%%%%%%%%%%%%%%%%%%%%%%%%%%%%
%\subsection{Spatial Reuse with per-Cell Multiround Delivery}
\subsection{Reuse with per-Cell Multiround Delivery}
\label{sec:frequency-reuse}

A reuse scheme of order $r$ consists of a coloring of the helpers with $r$ colors, such that helpers with the same color are associated to the same 
subband or time slot, and helpers with different colors corresponds to orthogonal subbands or time slots. 
In particular, two helpers $h, h'$ such that $\Bc(\theta_h,a_{\rm cell}) \cap \Bc(\theta_{h'},a_{\rm interf}) = \emptyset$
%, and obviously
%also $\Bc(\theta_h,a_{\rm interf}) \cap \Bc(\theta_{h'},a_{\rm cell}) = \emptyset$ 
can be associated to the same color and will never cause
interference conflicts, since the users that can receive from $h$ are not interfered by $h'$ and vice versa. 
%Hence, the minimum reuse distance in order to avoid conflict edges in the graph such as in Fig.~\ref{fig:scenario1} (b) 
%for any arbitrary user placement is $a_{\rm cell} + a_{\rm interf}$. 
However, this is just a sufficient condition. More in general, for a given helper and user placements $\Theta$ and $\Phi$  there exist a minimum $r \in [H]$ for which there exists a reuse scheme of order $r$ such that the network graph has no conflict edges.
The delivery strategy proposed in this section consists of two sub-problems: 1) for given $\Theta, \Phi$ find a reuse scheme of 
order $r$ with $r$ as small as possible avoiding all conflict edges; 2) for the given reuse scheme, associate users to helpers such as the system is feasible and the worst-case delivery time is minimized. 
	
We define an undirected conflict graph $G=(V,E)$, with $V$ and $E$ denoting vertex and edge sets, respectively, where vertices $V = [H]$ correspond to helpers and $E$ contains  edges $(i,j)$, for all $i,j \in V$, such that there is  a user at distance distance $ \leq a_{\rm interf}$ from both helpers $i$ and $j$. 
%In a  realization of network if  two helpers are neighbor but there is no user who is receiving simultaneously neither  signal nor interference, there is no edge between these two helpers in conflict graph.
The classical integer linear  programming model for vertex coloring problem is defined 
as follows: define the binary $0-1$ variables ${y_c : c \in [H]}$ and  ${x_{i,c} : (i, c) \in [H] \times [H]}$, with $y_c = 1$ if color $c$ is used in the reuse assignment and $x_{i,c} = 1$ if vertex $i$ is assigned color $c$.  Then, the minimization of the reuse order can be written as:

 \begin{subequations}
 	\label{eq:int_prog}
 	\begin{align}
 	\underset{}{\text{minimize}} \;\; \sum_{c=1}^{H} y_c & \\
 	\text{subject to:}  
 	&    \sum_{c=1}^{H}  x_{i,c} = 1,~~~ \forall i \in V \label{constraint1}\\
 	&   x_{i,c}+x_{j,c} \leq y_c, ~~~\forall (i,j) \in E, \forall c \in [H]  \label{constraint2}\\
 	&  x_{i,c} \in \{0,1\}, ~~~\forall i\in V,  \forall c \in [H]   \\
 	& y_c \in \{0,1\} ,~~~  \forall c \in [H] . 
 	\end{align}
 \end{subequations}
 It turns out that the minimum $r$ (solution of \eqref{eq:int_prog}) is the chromatic number of the graph $G(V,E)$, and problem \eqref{eq:int_prog} is known to be NP-hard for general graphs $G$. 
 Several important heuristic  approaches have been proposed to solve the vertex graph coloring problem. 
 One of the best known algorithms is \textsc{DSatur}  in \cite{DSATUR}. 
That has been shown to provide the optimal solution for bipartite graphs with running time of
 $O(|V|^2)$.
The pseudo-code of the algorithm can be found  in  \cite[Fig. 2.8]{lewis2015guide}.
%The vertex coloring  of aforementioned conflict graph is a NP-hard problem.

%%%%%%%%%%%%%%%%%%%%%%%%%%%%%%%%%%%%%%%%%%
After solving graph coloring subproblem, then we address the user-helper association problem. Assume that a reuse scheme of order $r$ has been found. 
%For example, for the network in Fig.~\ref{fig:scenario1} a reuse scheme with $r = 3$ eliminating the interference conflicts of corresponding graph.
%In the resulting colored graph, some users may be served by different helpers. For example, in Fig.~\ref{fig:scenario1} user 1 can be only associated to helper 1, 
%user 2 can be only associated to helper 2, and user 5 only to helper 3. However, we can assign user 3
%  either to the helper 2 or to helper 3, user 4 either to helper 3 or to helper 4, and user 6 either to helper 2 or to helper 4. 
%\begin{figure}
  %  \centerline{\includegraphics[width=7cm]{reuse.pdf}}
    %\caption{A reuse scheme with $r = 3$ colors eliminating the interference conflicts for the network of Fig.~\ref{fig:scenario1}.}
 %\label{fig:reuse}
%\end{figure}
In general, on the colored graph we need to find the optimal user-helper association that minimizes the worst-case delivery time. 
For a given association, the delivery  at each helper achieved by the multiround delivery scheme in Section~\ref{sub:direct extension}.  Let $\Pc_\ell^h$ denote the set of users in caching group $\ell$ associated to helper $h$.
The  multiround delivery load of helper $h$, denoted by $R_h$, is given by expression (\ref{eq:eliarate}) with occupancy numbers
$|\Pc_{[1]}^h| \geq |\Pc_{[2]}^h| \geq \cdots \geq |\Pc_{[L]}^h|$.
The corresponding delivery time over the fronthaul link is simply given by $R_h F/C_{\rm front}$ while the delivery time over the access (broadcast) downlink is given by  $r R_h F/C_{\rm access}$, where the factor $r$ comes from the fact that with reuse of order $r$, the access channel is used only 
for a fraction $1/r$ of the total access transmission resource. It follows that the worst-case delivery time for the reuse scheme with given user-helper association
is given by  
\begin{equation} \label{eq:T_reuse}
T^{\rm reuse} = \max \left \{ \frac{F}{C_{\rm front}}, \frac{ r F}{C_{\rm access}} \right \} \times  
\max_{h \in [H]}  \left \{ R_h \right \}. 
\end{equation} 

In order  to minimize $T^{\rm reuse}$,  define the binary variables $x_{h,k}$ such that $x_{h,k} = 1$ if user $k$ is associated to helper $h$ and $x_{h,k} = 0$ otherwise.  Denoting by $E$  the set of solid colored edges in the colored graph, we have that $x_{h,k} = 0$ for all $h,k$ for which there is no edge in $E$. 
The cardinality of the caching group $\Pc_\ell^h$ for a given association $\{x_{h,k}\}$ is given by 
\begin{equation} \label{card-ell-h}
|\Pc_\ell^h| = \sum_{k : (h,k) \in E, k \in \Pc_\ell} x_{h,k}.
\end{equation} 
For given non-negative integers $A_1, \ldots, A_L$, let the $[\cdot]$ subscript notation denote the sorting permutation such that
$A_{[1]} \geq A_{[2]} \geq \cdots \geq A_{[L]}$ and define the quantity 
\begin{equation}
\label{multiround-numbers}
R_{\rm multi}(A_{1}, \ldots, A_{L}) = \sum_{\ell \in [L - t^{\prime}]} \frac{A_{[\ell]} \binom{L-\ell}{t^{\prime}}}{\binom{L}{t^{\prime}}}. 
\end{equation}
Hence, taking  (\ref{card-ell-h}) into (\ref{eq:eliarate}) and using the definition in (\ref{multiround-numbers}) 
we arrive at the following optimal user-helper association problem
 \begin{subequations}
 \label{eq:opt_user-helper}
 \begin{align}
    \underset{\{x_{h,k} \in \{0,1\}\}}{\text{minimize}} \;\; \alpha & \\
\text{subject to:}  
    & \quad  R_{\rm multi}(A^h_{1}, \ldots, A^h_{L})  \leq \alpha,  ~\forall h \in [H],  \label{conb} \\
    & \quad  A^h_\ell  \geq \sum_{k : (h,k) \in E, k \in \Pc_\ell} x_{h,k}, ~\forall h \in [H], \;\; \ell \in [L], \\
    & \quad \sum_{h : (h,k) \in E} x_{h,k} \leq 1, \\
     & \quad x_{h,k} \in \{0,1\} \label{cone},\\
     & \quad   A^h_\ell \in \RR_+.
 \end{align}
 \end{subequations}
Since the function  $R_{\rm multi}(A^h_{1}, \ldots, A^h_{L})$ there is defined through a sorting permutation,  the optimization problem  \eqref{eq:opt_user-helper} is non-linear. 
A way out of this problem can be found by noticing that the coefficients  $\frac{\binom{L-\ell}{t^{\prime}}}{\binom{L}{t^{\prime}}}$ appearing in (\ref{multiround-numbers}) are decreasing with $\ell$, it follows that 
 for a given set of occupancy numbers $A_1, \ldots, A_L$, the permutation that sorts them in non-increasing order yields the maximum of the quantities
 \begin{equation}
\sum_{\ell \in [L - t^{\prime}]} \frac{A_{\pi_\ell} \binom{L-\ell}{t^{\prime}}}{\binom{L}{t^{\prime}}}
 \end{equation}
over any permutation $\pi$.
Hence, problem (24) can be linearized by replacing each constraint \eqref{conb} by the set of $L!$ constraints 
 \begin{equation} \label{replacement}
 \sum_{\ell \in [L - t^{\prime}]} \frac{A^h_{\pi_\ell} \binom{L-\ell}{t^{\prime}}}{\binom{L}{t^{\prime}}} \leq \alpha, \;\;\; \forall \; \pi \in \Pi_L,
 \end{equation}
where $\Pi_L$ denotes the set of all permutations of order $L$,  where only one constraint (corresponding to the sorting permutation [.]) actually bites.

 \paragraph*{Greedy user association algorithm} \label{sec:greedy}
 The drawback of the exact optimization user association method is that it generates $H \times L!$ constraints while in fact we need only $H$ of such constraints since  all the others are redundant. 
In order to reduce the complexity, we propose a  greedy approach where 
the complexity order of our greedy algorithms is upper bound   by $L \log(L) \times K $. 
We define the set of helpers who can transmit message to user $k$ as $\Hc_k = \{h: \phi_k \in  \Bc(\theta_h, a_{\rm cell})\}$. The proposed greedy user-helper association works as follows: i)
Initialize the groups $\Pc^h_\ell$  by including the users that have unique assignment, i.e., for which $|{\cal H}_k| = 1$; ii)
For all users k for which $|{\cal H}_k| > 1$,  associate them one by one, by selecting at each time the assignment that 
	yields the minimum increase in the total objective function $\max_{h \in [H]} R_h$.

\begin{example}  \label{ex:reuse}
	 Consider a network with $H=4$ cells and $K=6$ as shown in Fig. \ref{fig:scenario1} and assume $L=3$ cache configurations with $t'=1$ 
		\begin{align*}
		\Pc_1=\{1,2,5\}, \ \Pc_2=\{3,6\}, \ \Pc_3 = \{4\}.
		\end{align*}   
	    A reuse scheme with $r = 3$ eliminating the interference conflicts is given in Fig.~\ref{fig:scenario1} (b) . Helpers $ 1$ and $2$ are assigned to frequency band $1$, while helpers $ 2$ and $ 3$ are  assigned to frequency band $2$ and $3$, respectively. 
		In the resulting colored graph, some users may be served by different helpers.
	   Notice that the multiround delivery length in slots of
		duration $\frac{F}{C_{\rm access} {L \choose t'}}$ is given by \eqref{eq:eliarate}. 
		In the first step, user $1$ is assigned to helper $h=1$ and user $2$ to helper $h=2$ and user 5 helper $h=5$.
		 Since assigning user $3$ to helper $2$ or $3$ will increase the number transmission slots equally, user $3$ is randomly assigned to helper $2$. By assigning  user $4$ to helper $3$ the number of transmission slots of helper $3$ will be increased to $3$; on the other hand, assigning this user to helper $4$ will increase the number of transmission slots of helper $4$ to $2$. Then the greedy algorithm assigns user $4$ to helper $4$. Finally user $6$  can be assigned to either  helper $2$ or helper $4$. By assigning the user to helper $2$  the number of transmission slots of this helper will increase to $5$ while assigning this user to helper $4$ will increase the number transmission slot of helper $4$ to $3$. Therefore, the greedy algorithm assigns  user $6$ to helper $4$.  The resulting delivery time is
\begin{equation} \label{delivery-ex1}
T^{\rm reuse}  = \frac{3 F}{C_{\rm access}} \times \frac{3}{3} = \frac{3 F}{C_{\rm access}}.
\end{equation}

\hfill $\lozenge$
\end{example}

\subsection{Avalanche scheduling scheme}
\label{sec:avalanche}

In this section we propose a routing and scheduling strategy that embraces collisions,  and resolves them over multiple time slots, 
thus obtaining an overall better worst-case delivery time.  
The users are divided into two groups: non-interference  and interference users. The non-interference users are those in the non-interference service area of the helpers given by the union over $h \in [H]$ of unique coverage regions $\Bc(\theta_h, a_{\rm cell}) \setminus \cup_{h' \neq h} \Bc(\theta_{h'}, a_{\rm interf})$.
The second group contains users that are served by at least one
helper and are receiving interference from neighboring helpers given by the union over $h \in [H]$ of  coverage regions   $ \Bc(\theta_h, a_{\rm cell}) \bigcap \cup_{h' \neq h} \Bc(\theta_{h'}, a_{\rm interf})$. 
%As described already for the fractional reuse scheme, users are divided into two groups: non-interference, and interference users. 
In reuse scheme all users within a cell are served with reuse factor $r>1$, even non-interference  users, while the helpers can serve these users  by  all frequency resource without any collision. Our proposed avalanche scheme 
creates platform where  users can be served reuse factor $1$.
We let $\Pc^h_\ell$ denote the sets of non-interference users uniquely associated to helper $h$ and belonging to caching group $\ell$.
At any point in time, the {\em delivery list} of helper $h$ is the union $\Lc^h = \cup_{\ell \in [L]}  \Pc^h_\ell$. The main difference between the
scheme of this section, referred to as  {\it avalanche} scheme, and the  spatial reuse scheme in  Section~\ref{sec:frequency-reuse} is that in the avalanche scheme the sets $\Pc^h_\ell$ and therefore the delivery list of each helper is dynamically updated.
The avalanche scheme starts by  scheduling all the non-interference users simultaneously by running multiround delivery in parallel for all helpers. 
Since the multiround load depends on the 
occupancy numbers $\{|\Pc^h_{[\ell]}| : \ell \in [L]\}$ and these may differ over the helpers, at some point  
some helper $h$ finishes its multiround delivery before  the others. As soon as $h$ finishes, it stops transmitting, so that some users that are interfered 
by $h$ but can be served by some $h' \neq h$ become non-interference users and can be added to the delivery list of helper $h'$. 
Notice that some helper $h$ may stop temporarily to transmit since its delivery list is empty, but may restart when some users in its service area 
$\Bc(\theta_h, a_{\rm cell})$ that are interfered by some other helper become non-interference users when such helper stop its transmission. 
The process continue until all users are served. This algorithm is linear with number of edges in the bipartite graph represented in \ref{fig:scenario1}.
 For sparse graphs, as typically induced by the geometric coverage model considered here combined with some admission control protocol that limits the number of active users in each cell, the degree of each edge is upper bounded by some constant independent of $H$ and $K$. Hence, the avalanche algorithm has linear complexity in the number of users $K$. 
 Before giving the detailed description of the avalanche algorithm, we illustrate it through the following example.
  
\begin{example}\label{ex:comp}
	Consider again the network of Fig.  \ref{fig:scenario1} with  $L = 3$ cache configurations and $t’ = 1$ (same as in Example \ref{ex:reuse}). 

		After decentralized caching, assume that the realization of the cache configurations is:
		\begin{align*}
		\Pc_1=\{1,2,5\}, \ \Pc_2=\{3,6\}, \ \Pc_3 = \{4\}.
		\end{align*}   
		At the beginning, the only helper with non-empty delivery list is $h = 1$, with $\Pc_1^{(1)} = \{1\}$. All other sets are empty. Notice that the multiround delivery length in slots of
		duration $\frac{F}{C_{\rm access} {L \choose t'}}$ is given by  \cite{parrinello2018fundamental, jin2016} 
		\begin{align}
		\Delta_h = \sum_{r \in [L-t^{\prime}]} |\Pc_{[r]}^h| \binom{L-r}{t^{\prime}},
		\label{eq:Lh}
		\end{align}
		where,  we always assume (up to re-labeling of the caching groups) that at any point in time we have ordered occupancy numbers
		$|\Pc_{[1]}^h| \geq |\Pc_{[2]}^h| \geq \cdots \geq |\Pc_{[L]}^h|$. 
		Helper 1 finishes its delivery in ${3-1 \choose 1} = 2$ slots and stops transmission. In doing so, its frees from interference users 2 and 5. 
		At this point the updated delivery lists contain $\Pc^{(2)}_{[1]} = \{2\}$ for helper 2, and $\Pc^{(3)}_{[1]}  = \{5\}$ for helper 3. All other lists are empty. 
		Helper 2 and 3 can serve users 2 and 5, respectively, in ${3-1 \choose 1} = 2$ time slots and stop transmission. At this point  the delivery list of helper 4 contains
		$\Pc^{(4)}_{[2]} = \{6\}$ and $\Pc^{(4)}_{[3]} = \{4\}$. Users 4 and 6 belong to different caching groups and therefore can be served simultaneously by multicast coded messages.
		This requires ${3-1 \choose 1} + {3 - 2 \choose 1} = 3$ slots. Finally, user 3 can be served either by helper 2 or by helper 3, again in 2 times lots. 
		Eventually, the total delivery time is given by 
\begin{equation} \label{delivery-ex2}
T^{\rm avalanche}  = \frac{F}{C_{\rm access}} \times \frac{2 + 2 + 3 + 2}{{3 \choose 1}} =   \frac{3 F}{C_{\rm access}}. 
\end{equation}
Notice that in this example the avalanche and the reuse scheme (see Example \ref{ex:reuse}) achieve the same delivery time.
\hfill $\lozenge$
\end{example}

In order to give a general pseudo-code of the avalanche algorithm we need to introduce some notation. 
Consider the network bipartite graph $G$ as in Fig.~\ref{fig:scenario1} (b). For a given subset of helpers $\Hc'$ and users $\Uc'$, let
$G(\Hc',\Uc')$ denote the corresponding subgraph obtained by keeping only the edges $\{(h,k) : h \in \Hc', k \in \Uc'\}$. Let 
$\Ec_s(G)$ be the set of solid edges (i.e., carrying useful signal) and $\Ec_d(G)$ the set of dashed edges (i.e., carrying interference) in graph $G$. 
For a given assignment of the caching groups $\Pc_1, \ldots, \Pc_L$, and a subset $\Ac \subset [H]$ of helpers, 
let re-define the symbol $\Pc_\ell^h$ to denote the set of users $k \in \Pc_\ell$ such that $\exists! \; (h,k) \in \Ec_s(G(\Ac, \Uc))$, where we use the unique 
existence symbol $\exists!$ to indicate that there is no other edge $(h',k) \in \Ec_s(G(\Ac, \Uc)) \cup \Ec_d(G(\Ac, \Uc))$ in the graph. 
For what said before, defining the delivery list of helpers $h \in \Ac$ as $\Lc^h = \bigcup_{\ell = [L]} \Pc^h_\ell$, letting only
the helpers in $\Ac$ to transmit (i.e., be {\em active}), it is clear that each helper in $h \in \Ac$ can serve by multiround delivery 
the users in $\Lc^h$ without interference from other helpers. 
Notice that in the avalanche delivery every XOR packet has size (in bits) $F / {L \choose t'}$ and since the helpers broadcast
these packets on the access downlink, and must receive them from their fronthaul link, the transmission duration of a packet is
given by $T_{\rm slot} = \frac{F}{{L \choose t'} \min\{ C_{\rm access}, C_{\rm front}\}}$. 
For a given active helper set $\Ac$ with delivery lists $\{\Lc^h\}$, we define the delivery epochs as the integer multiples of $T_{\rm slot}$ at which 
each column of the delivery array $\Bm^h$ induced by $\Lc^h$ is finished. Explicitly, these are given by
\begin{equation} \label{epochs}
\Delta_{h,j_h} = \sum_{i = 1}^{j_h} \left ( {L \choose t'+1} - {L - |\Rc^h_{j_h}| ] \choose t'+1} \right ), 
\end{equation}
for $ j_h=|P^h_{[1]}|$ for each $h \in [H]$, representing the index of the last column  in the delivery array of helper $h$ and  as defined before, $L - |\Rc^h_{j_h}|$ denotes the number of zeros in column $j_h$ of $\Bm^h$. 
%for $j_h = 1, \ldots , |\Pc^h_{[1]}|$ where, as defined before, $L - |\Rc^h_{j_h}|$ denotes the number of zeros in column $j_h$ of $\Bm^h$. 
Notice that, given the cache group assignment $\Pc_1, \ldots, \Pc_L$ and the residual graph $G(\Ac, \Uc)$, 
the delivery lists $\Lc^h$ and arrays $\Bm^h$ therefore the sequence of delivery epochs $\{\Delta_{h,j_h}\}$ are uniquely determined. 
Then, the avalanche algorithm works as follows: 

{\bf Initialization:} Let $\Uc_{\rm unserved} = [K]$, $D = 0$, and $\Ac$ be a maximal set of helpers with non-empty delivery lists in 
$G(\Ac, \Uc_{\rm unserved})$ (e.g., this can be easily found in a greedy fashion). 

\begin{enumerate}
\item {\bf Find the set of helpers finishing the current served array column:} $\Fc = \{ h \in \Ac : \Delta_{h,{j_h}} - D > 0 \;\; \mbox{is minimal}\}$, and let $\Delta_{\min} =\min\{h\in \Ac:\Delta_{h,j_h}-D \}$  such minimal ``next epoch'' value. 
\item {\bf Increment time:} $D \leftarrow D + \Delta_{\min}$ (at this point, all helpers in $\Fc$ have finished delivery to the users in the $j_h$-th 
column of their delivery array, denoted collectively as $\partial \Uc = \{ \Bm^h_{:,j_h} : h \in \Fc\}$). 
\item {\bf Update unserved users:} $\Uc_{\rm unserved} \leftarrow  \Uc_{\rm unserved} \setminus \partial \Uc$. 
\item If $\Uc_{\rm unserved} = \emptyset$, exit. 
\item {\bf Identify stopping helpers:} for all $h \in \Ac$ such that $\Lc^h = \emptyset$, $\Ac \leftarrow \Ac \setminus \{h\}$.   
\item {\bf Update delivery arrays of active helpers:} for all $k \in \Uc_{\rm unserved}$ such that
$\exists! \; (h,k) \in \Ec_s(G(\Ac, \Uc_{\rm unserved}))$, add $k$ to $\Lc^h$ in the ``next available column'' of $\Bm^h$ and Update $j_h$ for each $h \in \Ac$. 
\item {\bf Reactivate helpers:} for all $h \in [H] \setminus \Ac$ such that $\exists! \; (h,k) \in \Ec_s(G(\Ac, \Uc_{\rm unserved}))$ for some 
$k$, add these users $k$ in $\Lc^h$ and increase $\Ac \leftarrow \Ac \cup \{h\}$. 
\item Go back to 1. 
\end{enumerate}
The ``next available column'' of $\Bm^h$ is the column $j_h$ of still completely unserved users,and for which the $\ell$-th group is empty
(i.e., it has a zero). Notice that it may be necessary to adjoin a new column to the right of the current array $\Bm^h$, i.e., 
the delivery arrays in general keep growing to the right, until there are no more users to add. 
Notice also that at each update of the delivery arrays, all future epochs successive to the current time $D$ must be updated accordingly, 
adding up future delivery intervals as in (\ref{epochs}). 
As said before, these are deterministic functions of $\{\Bm^h\}$ and therefore such updates are trivial. 
Under the condition that all users are at least one helper at distance $a_{\rm cell}$, the algorithm terminates with probability $1$ and
the resulting delivery time is $T_{\rm avalanche} = D \times T_{\rm slot}$.

%%%%%%%%%%%%%%%%%%%%%%%%%%%%%%%%%%%%%%%%%%%%%%%%%%%%%
\section{Results and discussions}
\label{sec:NumericalResults}

In this section we present some numerical results illustrating interesting features of the schemes proposed for the topological and the broadcast/collision network models defined in Section \ref{sub:system} and \ref{sub:system1}, respectively. 
For the topological network, given the helper placement $\Theta$ and user placement $\Phi$ in the plane, we associate 
user $k$ in position $\phi_k$ to all helpers $h$ such that $\phi_k \in \Bc(\theta_h,a_{\rm sig})$, for some signaling radius $a_{\rm sig }$. As said before, we are not concerned with coverage probability and only the users associated at least with one helper are considered, otherwise the delivery would be infeasible. 
For the broadcast/collision model, we produce the signal/interference graph by defining the radius $a_{\rm cell}$ and $a_{\rm interf}$ as described in Section \ref{sub:system1}. For the same placements $\Theta$ and $\Phi$ the two models are related. In particular, driven by practical considerations, we let $a_{\rm cell} \leq a_{\rm sig } \leq a_{\rm interf}$. In addition, we let $C_{\rm access}$ be equal in both cases, such that the sum rate of the (orthogonal) links outgoing from each helper in the topological model is equal to the broadcast downlink capacity of the helpers in the interference model. 
 
We consider the following parameters throughout the simulations: 
in the topological model $a_{\rm sig} = 220$m, while in the broadcast/collision model $a_{\rm cell} = 200$m and $a_{\rm interf}/a_{\rm cell}=1.2$.  
The placements $\Theta$ and $\Phi$ are generated according to independent homogeneous {\em Poisson Point Processes} (PPP) 
with densities $\lambda_h = 7$ helpers per km$^2$ and  $\lambda_u = 20 \times 7$ users per km$^2$, within a circle of area  $A = \pi \Rsf^2$ with radius $\Rsf = 1$km. This results in an average $H= 20$ helpers and $K = 400$ users in the whole area.  By fixing an integer $L < K$ and generating $L$ 
cache configurations as described in Section \ref{sec:grouping}, each user chooses one of the $L$ configurations randomly and independently.
Hence, the PPPs of users with the same cache configurations are independent thinnings (with thinning factor $1/L$) of the user placement PPP. 
The total delivery time is obtained by averaging over several realizations of the network and users cache configurations.
Without loss of generality, we let $F=1$ bit and $C_{\rm front}=1$ bit/s. 
Therefore, our ``normalized'' results
are easily translated in actual bit/s by choosing the ratio $F/C_{\rm front}$. 

\begin{figure}[t]
	\centering
	\includegraphics[width=0.55\textwidth]{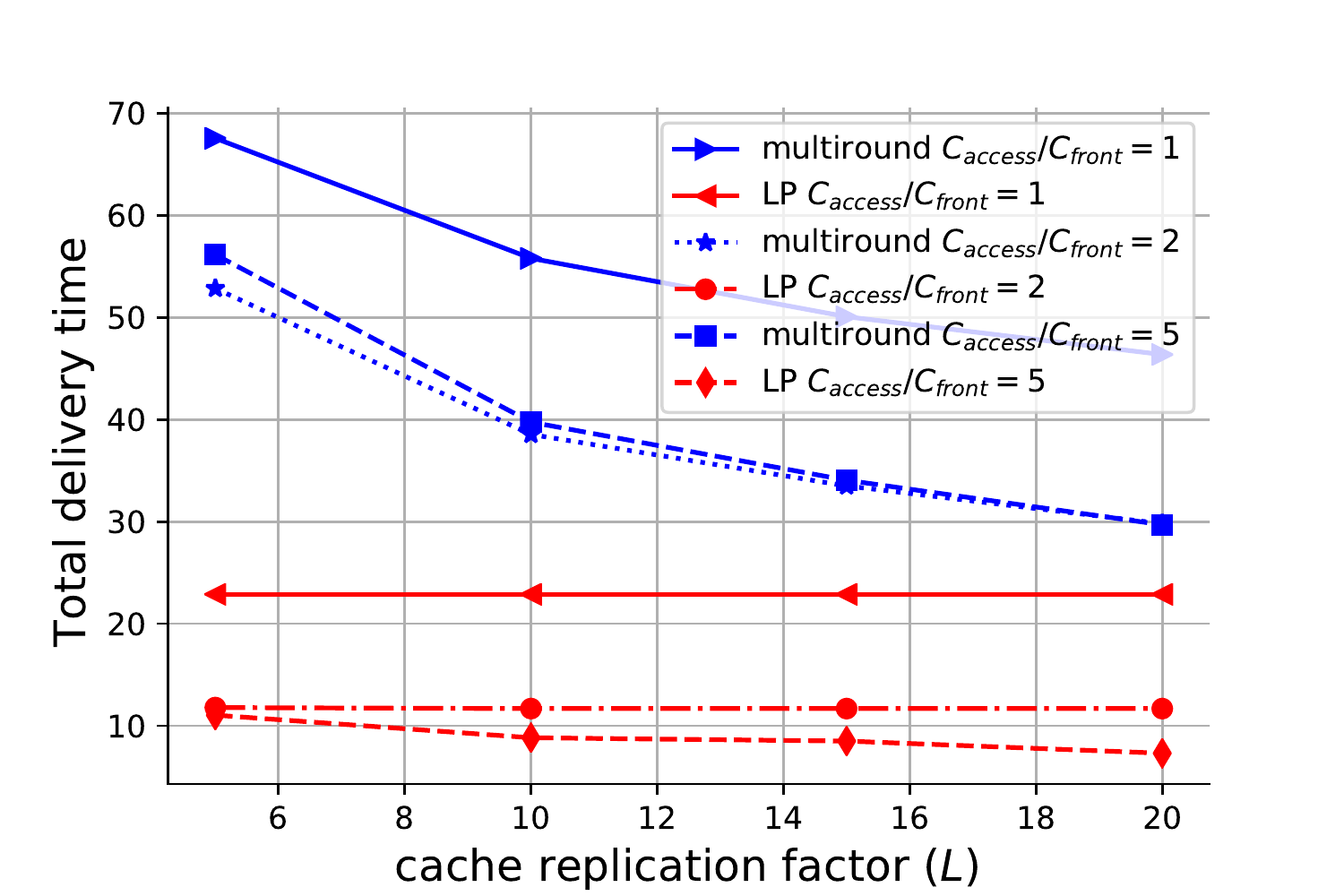}
	\caption{Delivery time versus cache replication factor with $\mu =0.2$ and $L= [5,10,15,20]$.   }
	\label{fig:cache_replication_L}
\end{figure}

%%%%%%%%%%%%%%%%%%%%%%%%%%%%%%%%%%%%%%%%%%%%
Fig.~\ref{fig:cache_replication_L}   compares the performance of the multiround delivery strategy 
(Section \ref{sub:direct extension}) and the newly proposed LP-optimized strategy (Section \ref{sub:novel decent method})  for the  topological  model. 
We notice that as 
the ratio $C_{\rm access}/C_{\rm front}$ increases, the worst-case delivery time of both schemes decreases but this improvement is limited
(for example, going from $C_{\rm access}/C_{\rm front} = 1$ to $C_{\rm access}/C_{\rm front} = 2$ the improvement is significant, but the improvement
from $C_{\rm access}/C_{\rm front} = 2$ to $C_{\rm access}/C_{\rm front} = 5$ is essentially negligible. This indicate the fact that for 
$C_{\rm access}/C_{\rm front} = 1$ the system bottleneck is the access segment, but for $C_{\rm access}/C_{\rm front} \geq 2$ the system bottleneck
is the fronthaul segment. 
Since the most interesting conclusions for the routing/scheduling algorithms are obtained when the system bottleneck is the access segment, 
in the next results we considered $C_{\rm access}/C_{\rm front} = 1$.\footnote{Of course, this is a qualitative consideration and for different network parameters the transition between these two bottleneck regimes
may appear at different ratio values.} 
Another interesting consideration is that while multiround delivery seems to be sensitive to the choice of $L$, 
the new LP optimized routing is very insensitive and reaches its minimum delivery time already for $L$ as small as $L = 5$. 
Since the subpacketization order depends (exponentially) on $L$, this makes the new scheme very attractive since
it can afford a very small value of $L$ with almost no degradation in performance. Overall, we notice that the new LP-based scheme significantly outperforms
the multiround delivery, although this was proved to be optimal in the case of a single helper (or, in our case, when each user can be served only by a single helper, such that the network graph becomes a tree).

%%%%%%%%%%%%%%%%%%%%%%%%%%%%5

\begin{figure}[t]
	\centering
	\includegraphics[width=0.55\textwidth]{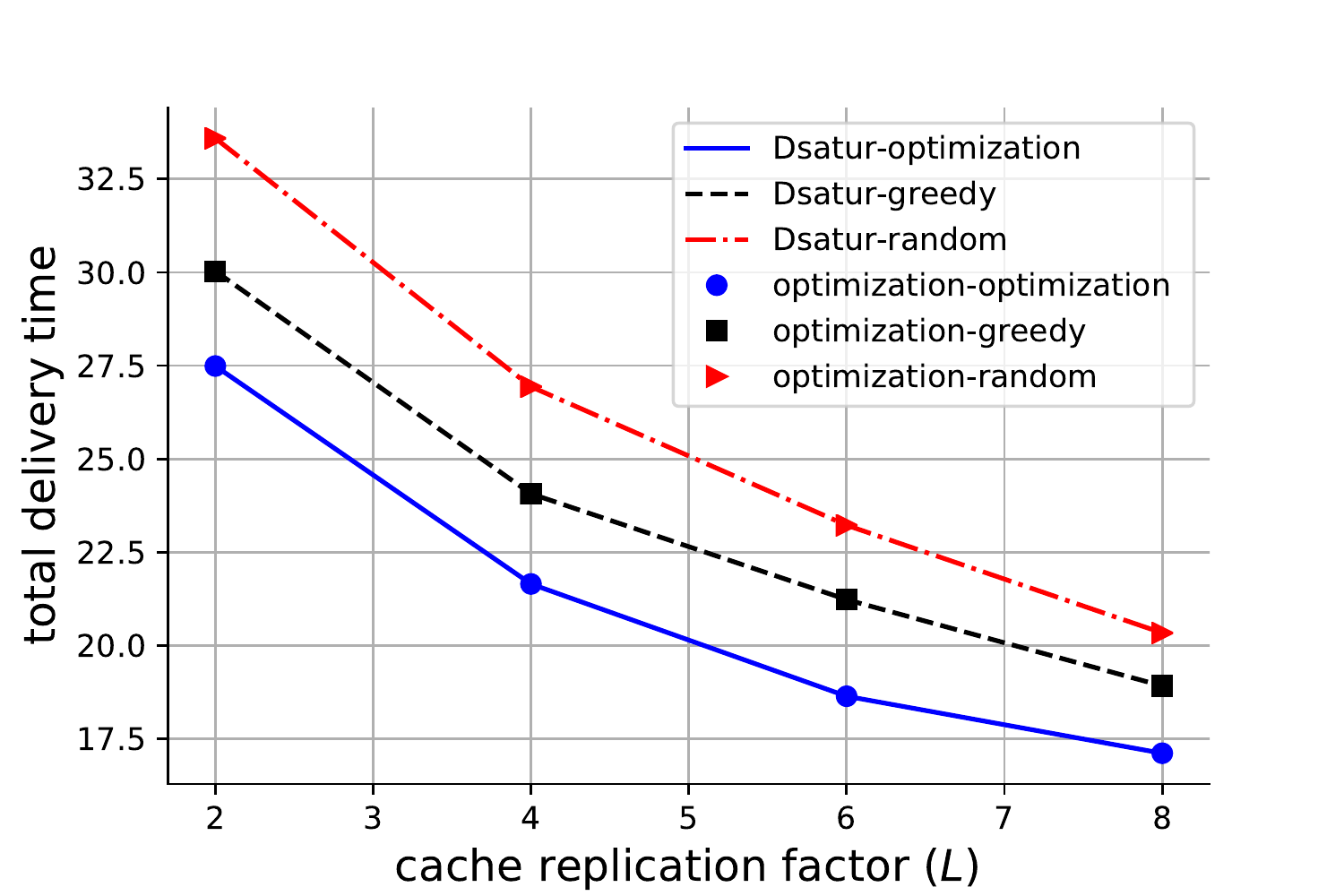}
	\caption{Delivery time versus cache replication factor with $\mu = 0.5$, $L= [2,4,6,8]$ and $C_{\rm access}/C_{\rm front} =1$ }.
	\label{fig:user_association}
\end{figure}

For the broadcast/collision model, Fig. ~\ref{fig:user_association} compares the performance for different user-helper association schemes
for the reuse scheme of Section \ref{sec:frequency-reuse}. The helper coloring is obtained by both solving the 
optimization problem in \eqref{eq:int_prog} and with the DSatur algorithm. As a mater of fact, even though the coloring optimization is NP-hard, for the network topologies considered here the problem can be easily solved by standard integer programming solvers for up to $H = 1000$ helpers. 
Furthermore, due to the sparse nature of the underlying graph, DSatur essentially provides optimal results. Hence, optimal or quasi-optimal coloring is really not a 
significant problem in most practical scenarios. After coloring, we consider user association according to the following three methods:
i) the relaxed LP user association in \eqref{eq:opt_user-helper}; ii) the greedy user association as described in Section \ref{sec:greedy}; iii)  
a random user association,  in which users are randomly assigned to one of the helpers at distance $a_{\rm cell}$.
The greedy scheme offers a significant improvement of the delivery time compared to the random association 
and has much lower computational complexity than solving the exact optimization in \eqref{eq:int_prog}. 

\begin{figure}[t]
	\centering
	\includegraphics[width=0.55\textwidth]{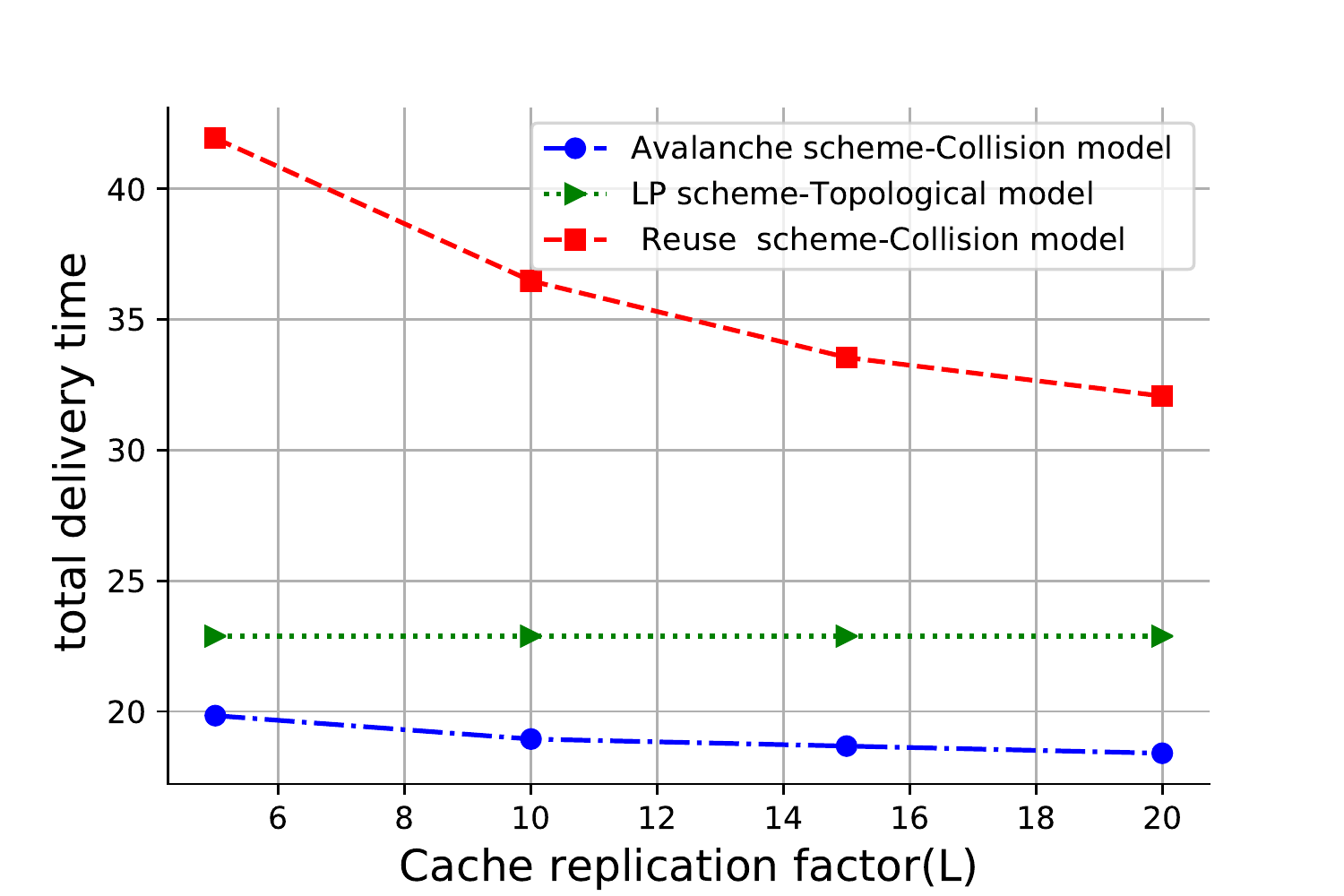}
	\caption{Delivery time versus cache replication factor with $\mu = 0.2$, $L= [5,10,15,20]$ and $C_{\rm access}/C_{\rm front} =1$.}
	\label{fig:avalanche}
\end{figure}

%%%%%%%%%%%%%%%%%%%%%%%
Fig. \ref{fig:avalanche} shows  a performance comparison between the reuse scheme
and the avalanche scheme for the broadcast/collision model with different parameters. 
The avalanche scheme is competitive since it exploits the multicast nature of the coded caching messages together with the network geometry. 
In addition,  the avalanche scheme is also very insensitive to the choice of $L$, such that we can choose $L$ as low as 5 and yet achieve nearly the best performance. This has an important impact on the subpacketization order. 

Fig. \ref{fig:avalanche} also shows the comparison between the broadcast/collision model and the topological model for the same value of
$C_{\rm access}$, i.e., the sum of the capacities of the links outgoing from each helper in the topological model is equal to the
helpers broadcast downlink rate of the broadcast/collision model. We notice that the broadcast/collision model with avalanche delivery
outperforms the topological model with LP-optimized delivery. This may be explained by the fact that
in the topological model each helper-user link is individual, such that the transmissions from helpers their served users is unicast.
In contrast, the broadcast nature of the helpers transmissions in the broadcast/collision model is better batched to the multicast nature of the coded caching messages. Evidently, the gain obtained by leveraging the broadcast nature of the downlink transmissions is enough to counter the loss due to 
collision interference. 

\section{Conclusion}

In this paper we investigated two variants of a general ``network-layer model''  formed by one server, several helper stations (APs/BSs) and several users, 
namely, a topological non-interference network with bounded sum capacity of outgoing links and a network with broadcast transmission from the helpers and collision model for interference.  For the first model, we proposed a novel routing optimization scheme  that outperform the direct extension 
of the so-called multiround delivery at each helper. The complexity of the novel method is linear in the the number of users.  
For the second model, we proposed a baseline scheme based on time/frequency reuse and user-helper assignment, 
and a new scheme nicknamed ``avalanche''  that is very reminiscent of CSMA-based random access. 
Our novel schemes have worst-case delivery that is very insensitive to the number of distinct cache configurations $L$. 
This has important consequences for the subpacketization order, since a small value of $L$ (yielding small subpacketization) 
can be chosen with minimal system performance degradation. Overall, our work essentially solves most of the problems outlined 
by previous literature as ``key issues'' for the application of coded caching in realistic scenarios. In particular, our schemes address
i) asynchronous user streaming sessions (by breaking large files in small blocks and using a low subpacketization order 
with minimal performance degradation and a completely decentralized prefetching); ii) scalability to large cellular-type networks with may users and helpers; 
iii) the HTTP end-to-end encrypted requests,  since the helpers are completely oblivious of the user requests and content files. 

\newpage

{\small
\bibliographystyle{IEEEtran}
\bibliography{references}
}

\end{document}